\documentclass[12pt]{article}
\input epsf.sty
\usepackage{wick,young}
\usepackage{amsfonts}
\newcommand{\beq}{\begin{equation}}\newcommand{\eeq}{\end{equation}}
\newcommand{\eqn}{\begin{eqnarray}}\newcommand{\enn}{\end{eqnarray}}

\def\CD{{\cal D}}

\def\CF{{\cal F}}

\def\CN{{\cal N}}
\def\CO{{\cal O}}

\def\CR{{\cal R}}

\newcommand{\eref}[1]{(\ref{#1})}

\newcommand{\Hu}{{\rm Hu}}
\newcommand{\Li}{{\rm Li}}

\newcommand{\schw}{Schwarzschild}

\newcommand{\ie}{{\em i.e.}\ }
\newcommand{\eg}{{\em e.g.},}

\newcommand{\AdS}[1]{{\rm AdS}_{#1}}

\newcommand{\bra}[1]{\langle{#1}|}
\newcommand{\ket}[1]{|{#1}\rangle}

\newcommand{\tr}[1]{{\rm tr}[{#1}]}

\newcommand{\BR}{\mathbb R}

\newcommand{\vx}{\vec{x}}
\newcommand{\vy}{\vec{y}}

\newcommand{\bbibitem}[1]{\bibitem{#1}\marginpar{#1}}

\newcommand{\ads}[1]{{\rm AdS}_{#1}}

\newcommand{\be}{\begin{equation}}
\newcommand{\ee}{\end{equation}}
\newcommand{\bea}{\begin{eqnarray}}
\newcommand{\eea}{\end{eqnarray}}

\def\Label#1{\label{#1}%
  \smash{\hbox to0pt{\raise1ex\hbox{\tiny[#1]}\hss}}}
\def\noLabels{\let\Label=\label}
\def\nobbibitem{\let\bbibitem=\bibitem}

\begin{document}

\renewcommand{\thepage}{\arabic{page}}
\setcounter{page}{1}
\noLabels % uncomment for final production
\nobbibitem % uncomment for final production

%\title
\rightline{\small hep-th/0508023} \rightline{\small UPR-1127-T}
\rightline{\small ITFA-2005-37} \rightline{\small DCTP-05/33}
\vskip 1cm \centerline{\Large \bf The Library of Babel:}
\centerline{\Large {\bf On the origin of gravitational
thermodynamics}} \vskip 1cm

\renewcommand{\thefootnote}{\fnsymbol{footnote}}
\centerline{{\bf Vijay Balasubramanian${}^{1,}$\footnote{vijay@physics.upenn.edu},
Jan de Boer${}^{2,}$\footnote{jdeboer@science.uva.nl},
Vishnu Jejjala${}^{3,}$\footnote{vishnu.jejjala@durham.ac.uk}, and
Joan Sim\'{o}n${}^{1,}$\footnote{jsimon@bokchoy.hep.upenn.edu}
}}
\vskip .5cm
\centerline{${}^1$\it David Rittenhouse Laboratories, University of Pennsylvania,}
\centerline{\it Philadelphia, PA 19104, U.S.A.}
\vskip .5cm
\centerline{${}^2$\it Instituut voor Theoretische Fysica,}
\centerline{\it Valckenierstraat 65, 1018XE Amsterdam, The Netherlands}
\vskip .5cm
\centerline{${}^3$\it Department of Mathematical Sciences, University of Durham,}
\centerline{\it South Road, Durham DH1 3LE, U.K.}

\setcounter{footnote}{0}
\renewcommand{\thefootnote}{\arabic{footnote}}

\begin{abstract}
We show that heavy pure states of gravity can appear to be mixed states to almost all probes.
For $\rm{AdS}_5$ Schwarzschild black holes, our arguments are made using the field theory dual to string theory in such spacetimes.   Our results follow from applying information theoretic notions to field theory operators capable of describing very heavy states in gravity.  For half-BPS states of the theory which are incipient black holes, our account is exact:  typical microstates are described in gravity by a spacetime ``foam'', the precise details of which are almost invisible to almost all probes.    We show that universal low-energy effective description of a foam of given global charges is via certain singular spacetime geometries.  When one of the specified charges is the number of D-branes, the effective singular geometry is the half-BPS ``superstar''.   We propose this as the general mechanism by which the effective thermodynamic character of gravity emerges.
\end{abstract}

\newpage

\tableofcontents

\newpage

{\small
\begin{quotation}
{\em There are twenty-five orthographic symbols. That discovery enabled
mankind, three hundred years ago, to formulate a general theory of the
Library and thereby satisfactorily solve the riddle that no conjecture
had been able to divine --- the formless and chaotic nature of virtually
all books....
The Library is total and its shelves register all the possible combinations of the twenty-odd
orthographic symbols (a number which, though extremely vast, is not infinite).} \\
%{\em Infidels claim that the rule in the Library is not ``sense,'' but ``non-sense,''
%and that ``rationality'' (even humble, pure coherence) is an almost miraculous exception.} \\
\rightline{--- Jorge Luis Borges, in ``The Library of Babel'' $\;\;\;\;\;\;\;\;\;\;\;\;\;\;\;\;\;\;\;\;\;\;$}
\end{quotation}}

\section{Introduction}

The microcanonical accounting of the entropy of black holes posits an enormous number of degenerate microscopic states.
In terms of these states, the information loss paradox may be evaded by showing that a pure initial state collapses to a particular pure black hole microstate whose exact structure can be deduced by suitably subtle measurements.
Given the importance of the problem, it is crucial to ask:
What do pure microstates look like and what sorts of measurements can distinguish them from each other?
Here we report that in string theory almost no probes are able to differentiate the microstates of a black hole.
Thus, if spacetime is placed in a very heavy pure state, it will appear mixed --- {\em i.e.}\ like a black hole --- to almost all finite precision measurements.
This explains how the existence of pure underlying microstates and the absence of fundamental information loss are consistent with the semiclassical observation of a thermodynamic character of black holes.

In Section 2, we analyze so-called large Schwarzschild black holes in AdS spacetimes.
These black holes have horizon size bigger than the scale set by the AdS curvature.
They are known to come into equilibrium with their thermal radiation because AdS geometries create an effective confining potential \cite{hawkingpage}.
Thus such black holes are stable, and we can ask how the underlying states that give rise to the large entropy can be identified via quantum mechanical probes.
We will examine black holes in five dimensions because string theory on $\AdS{5}\times S^5$ admits a dual description in terms of a superconformal $SU(N)$ gauge field theory propagating on a three-sphere \cite{ads}.
Here $N$ is related to the AdS curvature scale $1/\ell$ via the string coupling $g_s$ and the string length $\ell_s$ as $\ell^4 = 2\pi g_s N \ell_s^4$.
When the length scale $\ell$ is large, which we require, so too is $N$.
In the dual field theory, we characterize the structure of operators of conformal dimension $O(N^2)$ that create black hole microstates.
Almost all such operators belong to a ``typical set'', which is characterized by the statistical randomness of the polynomial in elementary fields that defines the operator.
Sanov's theorem from information theory shows that deviations from typicality are exponentially suppressed in the conformal dimension.
Atypical operators are vanishingly improbable in the large-$N$ limit.
We then argue that almost no probe correlation functions can distinguish two typical states from each other.
We also explore what questions about black holes are easy to answer using probe measurements.

Our arguments about AdS-Schwarzschild black holes are made in the dual
field theory because the underlying non-supersymmetric microstates
have not been given a geometric description. For this reason, in
Section 3 we examine half-BPS states of the theory which have been
explicitly described in both the AdS spacetime and the dual field theory.
Very heavy half-BPS states are incipient black holes --- a small
amount of energy beyond the BPS limit is expected to create a horizon.
In field theory, these states are described in terms of free fermions
in a harmonic potential, or equivalently, in terms of Young diagrams
enumerating gauge-invariant operators \cite{CJR,Berenstein}.
Like the \schw\ case, almost all states can be characterized by the
statistical randomness of the operators that make them. Specifically,
almost all such states are associated to Young diagrams whose shapes
are small fluctuations around a certain limiting curve. This
characterization of typical states is done in two different
statistical ensembles: with and without a restriction on the number of
columns in the Young diagrams. Physically, the second ensemble
corresponds to fixing the number of D-branes in the state.
The infinite temperature limit in the second ensemble
describes the set of typical states underlying the superstar \cite{myers}.

In Section 4 we relate the exact quantum mechanical description of
the microstates, and their averages studied in Section 3 with
their semiclassical gravitational description.\footnote{Our work is
related to the proposals of Mathur and others \cite{Mathur,omm,mathura,mathur,globalads3,simonvishnu,bena} in the context of the microstates of the D1-D5 system.   These works demonstrated the existence of non-singular geometries without horizons matching the quantum numbers of certain D1-D5 microstates.     The work of \cite{llm} can be regarded as providing the analogous half-BPS geometries in $\ads{5} \times S^5$.    Our perspective on these results is that they provide a classical configuration space that requires quantization.   We are using the field theory dual to $\ads{5} \times S^5$ to carry out this quantization and carefully study the semiclassical limit. }  We develop techniques for characterizing properties of
typical states in terms of phase space distributions and then
establish a dictionary with the classical
asymptotically AdS geometries to which they correspond. We
establish the criteria for half-BPS states to have such
semiclassical descriptions and show that a typical very heavy
state corresponds to a spacetime ``foam''.   The precise structure of the foam, and whether it looks more quantum or classical, depends on the observable.   We argue that almost no semiclassical probes will be able to distinguish different foam states, and that the resulting
effective description gives a singular geometry. For spacetimes in
which the overall mass and charge are specified the effective
geometry is a new singular configuration that we call the
``hyperstar'' (see also \cite{buchel}). For spacetimes in which
the number of D-branes is additionally specified, the effective
geometry is again singular, giving rise in the maximum entropy limit
to the well-known ``superstar''  \cite{myers} geometry.

%This curve describes the universal part (up to exponentially small deviations) of correlation functions computed in the typical states.

In Section 5, we argue for the universality of a subset of observables
in the one-half BPS sector. We also discuss particular
observables which are more efficient at discriminating 
between different microstates, but argue against the feasibility of
constructing such probes.

This universal low-energy effective description of underlying smooth quantum
states is the origin of the thermodynamic character of heavy gravitating systems.

\section{Schwarzschild black holes}

Consider the metric for a Schwarzschild black hole in $\AdS{5}$:
\be
ds^2 = -\left[ 1 + \frac{r^2}{\ell^2} - \frac{r_0^2}{r^2} \right] dt^2 +
        \left[ 1 + \frac{r^2}{\ell^2} - \frac{r_0^2}{r^2} \right]^{-1} dr^2 + r^2 d\Omega^2 + d(S^5(\ell)),
       \Label{bholemet}
\ee
where $\ell$ is the radius of $\AdS{5}$ and of $S^5$.
The horizon is at
\be
r_h \simeq \frac{\ell}{\sqrt2} \left[ -1 + \left( 1 + \frac{4r_0^2}{\ell^2} \right)^{1/2} \right]^{1/2}.
\ee
The difference between mass of the black hole and the mass of empty $\ads{5}$ is
\be
M = {3\pi r_0^2 \over 8G_5}.
\ee
Dropping numerical constants, the five-dimensional Newton constant is
$G_5 \simeq g_s^2 \alpha^{\prime 4}/\ell^5$.
Also recall the $\AdS{}$/CFT dictionary relating string theory on $\ads{5}$ to $\CN = 4$, $SU(N)$ super-Yang-Mills theory at large-$N$:
\bea
&& g_s N \simeq g_{YM}^2 N \equiv \lambda = {\rm fixed\ 't\ Hooft\ coupling},  \nonumber \\
&& \ell \simeq \ell_s (g_s N)^{1/4}, \\
&& M\ell \simeq \Delta = {\rm conformal\ dimension\ of \ the\ dual\ operator}. \nonumber
\eea
For a large black hole, $r_h \simeq \ell$ which implies $r_0 \simeq \ell$.
(For a small black hole, $r_0 \simeq r_h \ll \ell$; 
such small black holes are unstable to localization on the sphere 
and we will disregard them here.) If a spacetime of this mass is 
described by a pure state in the CFT, the conformal dimension of 
the operator making the state is
\be
\Delta = M\ell \simeq \frac{\ell^3}{G_5} \simeq \frac{\ell^8}{g_s^2 \alpha^{\prime 4}} \simeq \frac{g_s^2 \ell_s^8 N^2}{g_s^2 \ell_s^8} \simeq N^2,
\Label{confdim}
\ee
which is independent of the Yang-Mills coupling.
In view of the fact that these states are not supersymmetric, and should renormalize, it is remarkable that (\ref{confdim}) is independent of the Yang-Mills coupling.

From the CFT point of view, a pure underlying black hole microstate is described as
\bea
\ket{\rm microstate} = \CO\ket0; && \Delta(\CO) \simeq N^2.
\eea
It is worth comparing this to the conformal dimensions of various more familiar states in the AdS/CFT correspondence: \\
\begin{center}
\begin{tabular}{|c|c|}
\hline
Supergravity states in $\AdS{d+1}$ & $\Delta \simeq \frac{d}{2} + \sqrt{\frac{d^2}{4} + M^2\ell^2} \simeq \CO(1)$. \\
\hline
Small strings & $E \simeq 1/\ell_s \longrightarrow \Delta \simeq \ell/\ell_s \simeq (g_s N)^{1/4} \simeq \lambda^{1/4}$. \\
\hline
Small black holes (unstable) &
$M \simeq \frac{\ell^5 r_0^2}{g_s^2 \alpha^{\prime 4}} \longrightarrow \Delta \simeq \frac{r_0^2 N^2}{\sqrt{\lambda \ell_s^4}}\simeq N^2 \frac{r_0^2}{\ell^2}$. \\
\hline
 D-brane states (giant gravitons) & $\Delta \simeq N$. \\
\hline
\end{tabular}
\end{center}
\noindent
Large black holes are extremely heavy states, with a conformal dimension that grows in the large-$N$ limit like $N$ D-branes.
The black hole entropy is
\be
S \simeq \frac{r_h^3}{G_5} \simeq \frac{\ell^3}{G_5} \simeq {\ell^8}{G_{10}} \simeq N^2 \simeq \Delta,
\ee
so we expect that the field theory dual to $\ads{5}$ space has $O(e^{N^2})$ operators of conformal dimension $\Delta = N^2$ that are neutral under the global symmetries of the Yang-Mills theory.

\subsection{Typical states of Schwarzschild black holes}

The fields of $\CN=4$ SYM are the gauge fields $A_\mu$, three complex adjoint scalars $X, Y, Z$, and four complex Weyl fermions $\psi_\alpha$.
Gauge invariant operators can be constructed as products of traces of polynomials built out of the fields of the theory and their derivatives, with Lorentz and gauge indices appropriately contracted.
We wish to construct operators of dimension $N^2$ that are not charged under either the $SO(4)$ rotation group or the $SO(6)$ R-symmetry.
An example gauge invariant operator constructed from the scalars is
\be
\CO = (\tr{X X^\dagger Y Y^\dagger Z}) (\tr{X^\dagger Y^\dagger Y^\dagger Y Z^\dagger X X Y}) (\tr{X X^\dagger
X^\dagger}) (\tr{Y^\dagger Y Y^\dagger Z Y Z^\dagger X^\dagger}) \ldots.
\Label{egoperator}
\ee
Derivatives could be included also as well as fermions and gauge fields, but we ignore these for the moment, since our arguments will not depend on the details of particular operators.

The operators in question are not BPS since the Schwarzschild black
hole breaks supersymmetry; thus terms like $X X^\dagger$ may appear.
(Since R-charges effectively count differences in say the number of $X$s and $X^\dagger$s, a chargeless operator should have equal numbers of a field and its conjugate.)
In the absence of supersymmetry, operator dimensions renormalize, and thus the classical dimension could be different from the full quantum corrected dimension $\Delta = N^2$ that we are seeking.
However, it is possible that for such heavy operators renormalization is minimal.
One can argue for this in three steps:
\begin{enumerate}
\item The conformal dimensional $\Delta = N^2$ is independent of the coupling.
Since this is derived from the mass of the dual black hole in the semiclassical limit, we learn that at least at large 't Hooft coupling the conformal dimension does not receive corrections as the coupling is changed.
This leaves open the possibility that the renormalization of such operators is large at weak coupling.
\item Large black holes in AdS, such as the ones we study, are usually modeled as thermal ensembles in the dual field theory.
It is known that the thermal free energy only renormalizes by a factor of $3/4$ from weak to strong \cite{threequarters}.  This suggests that the energies of the heavy states entering the thermal ensemble also do not renormalize substantially from weak to strong coupling.
\item A possible explanation for this lack of renormalization despite the absence of supersymmetry is that such enormously heavy states are effectively classical because of cancellations between the contributions to the self-energy from effectively random interactions between various pieces of large operators like (\ref{egoperator}).
\end{enumerate}
In view of these arguments we will start in free field theory and ask how to construct neutral operators of large conformal dimension.
In any case, we will see that the basic result, that almost no probes can identify a state created by a very large operator, will not depend very much on the dimension being precisely $N^2$, and that interactions will not modify our conclusions.

At large-$N$, low dimension operators that are polynomials in traces are approximately orthogonal in the sense that the states they create have small overlap.
However, at large conformal dimension the trace basis mixes heavily even in the free field limit.
For example, an orthogonal basis of ``giant graviton'' D-brane operators of dimension $\sim N$  is constructed from determinants and sub-determinants of the scalar fields \cite{states,holey,BHLN,BBFH}, even though these can of course be expanded as linear combinations of traces.  
In addition, in an $SU(N)$ theory a trace of a product of more than $N$ fields decomposes into a sum of smaller traces.
We will not discuss the problem of finding an orthogonal basis of non-supersymmetric operators of dimension $O(N^2)$ because this will not be necessary.
We will argue that almost all operators of high dimensions have the properties we seek, and thus, in free field theory, any linear combination of them will do.
Subject to these caveats, which we will return to later, each operator can be thought of as a sentence built of words (traces), each of which is a string of letters in the alphabet provided by the elementary fields of the theory and their derivatives.
Black hole microstates are created by operators with very large dimensions ($\Delta \sim N^2$) and hence (ignoring traces of gauge indices) involve polynomials of length of order $N^2$.
Theorems in information theory characterize the structure of such polynomials.
As the conformal dimension becomes large, almost all operators will belong to a ``typical set'' characterized by the statistical randomness of the sequence of fields in the polynomial.
Specifically, for operators in the typical set the probability that a randomly chosen letter in the polynomial is $x$ is given by the uniform distribution $q(x) = 1/k$, where $k$ is the size of the alphabet.  (For the moment we are ignoring derivatives as ``letters in the alphabet''.)
In particular, this ensures that for operators in the typical set, the number of $X$s will be the same as the number of $X^\dagger$s, etc.\ so the R-charges will automatically vanish.
Deviations from typicality are exponentially suppressed.
For example, the likelihood that a randomly chosen operator of length $\Delta$ has a letter distribution $p(x)$ is given by
\begin{equation}
P(p(x)) = 2^{-\Delta D(p||q)}
\Label{sanov}
\end{equation}
where $D(p||q)$ is the Kullback-Liebler distance\footnote{Calling this a {\em distance} is a slight abuse of terminology.
$D(p||q)$ is not symmetric and does not satisfy the triangle inequality, and so is not a metric distance.}
or relative entropy between probability distributions
\begin{equation}
D(p||q) = \sum_i p(i) \log_2 \left( {p(i) \over q(i)} \right).
\Label{kkdist}
\end{equation}
This is an approximation at large $\Delta$ of the exact answer
\be
P(p(x)) = \left( \begin{array}{c} \Delta \\ \Delta p(1) \,\, \Delta p(2) \,\, \ldots \end{array} \right)
q(1)^{\Delta p(1)} q(2)^{\Delta p(2)} \ldots
\ee

In the present case, the precise result requires some modification to include the distribution of traces and derivatives, and the fact that the different Yang-Mills fields have different conformal weights.
For example, higher derivatives are permitted, thus enlarging the ``alphabet'', although equations of motion will relate many of these operators.
A proper accounting of these effects is necessary in order to obtain $e^{N^2}$ polynomials of length $N^2$ that are necessary to give the black hole entropy. This counting was studied in detail in \cite{ent1,ent2,ent3}\footnote{In
\cite{ent2} abelian theories are studied, whereas \cite{ent1,ent3} consider the nonabelian situation.}
and by Legendre transforming the result for the free energy at high temperature and zero coupling we obtain the
asymptotic density of states
\be
N(\Delta) \simeq e^{\frac{4}{3} \pi N^{\frac{1}{2}} \Delta^{\frac{3}{4}}}.
\ee
For conformal weights $\Delta\sim N^2$ this indeed yields the right $e^{N^2}$ degeneracy of operators.

Roughly speaking, the typical black hole microstate operator will look like a random polynomial with traces randomly distributed in it.
Operators such as
$$
\CO' = (\tr{XYXYXYXYXYXY
\ldots})(\tr{XYXYXYXYXYXY
\ldots})\ldots
$$
with discernible order are exponentially rare at large conformal dimension.
To gain some intuition for these statements consider single trace operators of dimension $n$ built out of $k$ species of scalar fields each having conformal dimension 1.
Then the distribution of letters in the typical state is $q(i) = 1/k$.
A randomly chosen operator will have a particular letter distribution $p(i) = 1/k + \delta_i/n$.
(The $\delta_i$ satisfy the constraint $\sum_i \delta_i = 0$ since the word length is fixed.)
Then Sanov's theorem \cite{CT} measures the probability of having a
large deviation from the expected distribution $q(x)$:\footnote{The
reason why this probability is not proportional to the length of the
random polynomial is the normalization chosen for the fluctuations
$\delta_i$. If we would have parametrized them as $p(i)=1/k +
\hat{\delta}_i$, this probability would indeed be manifestly
suppressed by the length of the operator $n$, that is, by its
conformal dimension $\Delta$.}
\be
P(p(x)) = 2^{-n D(p||q)} \simeq \exp\left( -\frac{k}{2n} \sum_i \delta_i^2 \right).
\ee
Large deviations are therefore exponentially suppressed.
We are simply describing a generalization of the central limit theorem ---
there are $O(\sqrt{n/k})$ variations in the frequencies of different fields appearing in the long operator.
This may appear large, but of course the ratio of the standard deviation to the mean,  $1/\sqrt{n/k}$, decreases as $n$ increases.

\subsection{Probing a typical state}

Consider states $\ket\CO = \CO\ket0$ created by operators of very high conformal dimension and a collection of probe operators $\CO_p$.

\vspace{0.2in}
\noindent {\bf Claim:}
Correlation functions of probe operators $\CO_p$ computed in very heavy states $\ket\CO$
({\em i.e.}\ $\bra\CO \CO_p \cdots \CO_p \ket\CO$)
depend only on the conformal dimension $\Delta$ and the global charges of $\CO_p$ and $\CO$, up to corrections of $O(e^{-\Delta})$, for almost all states $\ket\CO$ and probes $\CO_p$.
\vspace{0.2in}

In Section 5, we will explore this claim for half-BPS states after
showing in Sections 3 and 4 that the resulting effective
description of underlying smooth quantum states in low-energy gravity
can be a singular spacetime. Here, in view of the challenges of
rigorously enumerating the microstates of a Schwarzschild black hole
we will content ourselves with an argument rather than a proof.

For purposes of argument, consider the two point correlation function of a light probe $\CO_p = \tr{XX}$ in a heavy state $\CO = \tr{XYZXXYYZ \cdots}$.
In free field theory, Feynman diagrams contributing to the correlation function are computed via Wick contractions of fields in $\CO_p$ and $\CO$, {\em e.g.},
\begin{equation}
\wick{21}{\bra0 \ {\rm tr}[ XYZ <*X <1X YYX \cdots]^\dagger \ {\rm tr}[XX]^\dagger \ {\rm tr}[>1X>*X] \  {\rm tr}[XYZXXYYX \cdots] \ \ket0}
\end{equation}
In free field theory each pattern of such Wick contractions makes a separate contribution to the overall correlation function.
Contractions that are planar give contributions that have the leading order in $N$ and require the probe operator $\CO_p$ to appear as a sub-word within the state operator $\CO$ as the minimum number of index lines are broken in this way.
The contribution from planar diagrams is thus controlled by the number of times the probe appears within the state, a quantity that is completely determined by the statistics of long random polynomials.
In fact, at each order in the $1/N$ expansion the contribution of different patterns of Wick contractions is determined by how often each of these patterns occur.
This is also controlled by the statistics of random polynomials.
Adding perturbative interactions does not change the argument --- one must then include the interaction vertices, but the number of times each Feynman diagram pattern occurs is still determined by statistics.
In effect the long polynomials making the state are creating a random ``background'' with which the probe interacts.\footnote{
This is reminiscent of computations of open string interactions in the presence of D-branes in Yang-Mills theory \cite{holey,BHLN,BBFH}.}
The resulting statistical character of the computation guarantees that correlation functions in any typical state will take a universal value depending largely on the charges that control the overall frequencies of patterns in the state operator.
Deviations from this universal value will be exponentially suppressed as they will depend on the tiny structural differences between specific heavy operators and the idealized typical state.

The argument above was made in the context of single trace operators made from scalar fields, but it is clear that increasing the alphabet to include all the fields of the Yang-Mills theory, derivatives and randomly sprinkled traces will not change the essential statistical character of the computation.   (Of course, the precise details of the Sanov theorem (\ref{sanov}) will be modified since multiple derivatives will give rise to an alphabet size that grows without bound.)  Neither will the computation of higher point correlators. The mixing that is expected between multi-trace operators of very high dimension only improves the argument for the following reason.   We should expect all operators in the trace basis that have given quantum numbers to mix with about equal strength.
Thus, in effect, the mixing will average correlation functions over the typical set of large dimension operators, further emphasizing the universality of the correlation functions.
Interestingly, following (\ref{sanov}), deviations from universality in the correlations should be controlled by the small parameter
\begin{equation}
e^{-\Delta } = e^{-N^2}
\end{equation}
Since the entropy of the Schwarzschild black hole of mass $\Delta$ is $S \sim N^2$, this suppression is exactly as expected for a large statistical system with $e^S$ underlying microstates.

Since our arguments for universality of correlation functions will
work for any sufficiently heavy state, why are we describing a
black hole and not just a thermal gas? In a conventional thermal
gas, an individual probe with energies much larger than the
thermal scale passes through the system unaffected. By contrast,
heavier probes of black holes are increasingly likely to be
absorbed.   Indeed, a
high conformal dimension probe does not decouple from a heavy
state. Rather, the many fields entering into the operator will
individually interact with the fields in the state operator,
increasing the effective strength of the interaction. This is
simply telling us that the field theory knows about gravity, and
that the effective coupling involves the energy of states.   The
standard field theory statement of decoupling, that high energy
probes decouple from a low temperature thermal bath, applies to
individual probes of very high momentum passing through a
long-wavelength thermal background. This is a different notion of
high energy than the one relevant here, namely the conformal
dimension of a very heavy probe.

To actually see the precise difference between a black hole and a
thermal state such as a neutron star,  and in particular the
presence of a classical horizon, it will be necessary to
investigate the behavior of the correlator as a function of
Yang-Mills coupling. The reason for this is that any conventional
star will collapse to become a black hole if the Newton constant
is sufficiently large. Thus we should expect that as the string
coupling (which is dual to the Yang-Mills coupling) is increased,
standard thermal states can become black holes. While some sort of
``phase transition'' at finite coupling is presumably involved, it
is possible that some of the classic properties of black holes,
such as the difficulty of extracting information can already be
understood as described above.    This is because the universality
of correlation functions is being governed by the randomness of
the polynomials in fields that make up an operator along with the
randomized distribution of traces that will produce an ensemble of
small and large words in the complete operator.   Indeed, at least
in perturbation theory at finite coupling, it is clear that the
statistical character of the arguments that we have made does not
change.   Perhaps all of this should be read as saying the
difficulty of identifying the underlying state of a complex system
is {\it not} by itself the signature of the presence of black hole
with an ``interior region'' that is causally disconnected from infinity in the classical limit.

%%%%%%%%%%% askable questions  %%%%

\subsection{Askable questions}

Given the statistical nature of the operators $\CO$ creating black hole microstates, how much information can different kinds of probes give us about this state?
Recall that the conformal dimension of $\CO$ is $\Delta \sim N^2$.
As argued above, the correlation functions of a probe can be used to piece apart the patterns of fields appearing within the definition of $\CO$.
Supergravity probes with $\Delta \simeq O(1)$ will be completely ineffective at this task for several reasons.
First, as we have argued, their correlators will be universal up to corrections of order $e^{-N^2}$.
Of course, one might compute the correlators of many different probes to separate the different microstates.
However, simple counting shows that there are insufficient light operators of dimension $O(1)$ to distinguish the $e^{N^2}$ expected microstates.
Stringy probes ($\Delta \simeq (g_s N)^{1/4}$) will not be much better since the number of possible patterns of letters within microstates of dimension $\sim N^2$ is so large.

Brane probes ($\Delta \simeq N$) will be better placed to disentangle the microstate.
In particular the long stretches of fields appearing in the definition of the brane operator might  or might not match sequences of fields in the microstate and these two conditions will lead to somewhat different correlation function responses.
One might also use other black hole microstates ({\em i.e.}\ other operators of dimension $O(N^2)$) as probes.
The correlation functions of almost all such probes would involve the interaction of a random polynomial (the probe) with a different random polynomial (the state), leading to a correlation function of universal form.
However, if the experimenter happens to choose a probe that is identical to the black hole microstate, a huge response will result.
Using black hole microstates as probes in this way is difficult because there should be $e^{N^2}$ such operators.
Thus we see that it is far easier to detect what the state of the black hole is {\it not} rather than what it is.\footnote{If the underlying theory is actually integrable, there will be an infinite number of possibly nonlocal conserved quantities.   Measuring these in detail would in principle identify the black hole state.   However, even in this case, one would expect to have to make a very large number of very precise measurements to distinguish among microstates.   From this perspective, our argument is simply that typical very heavy states are very similar to each other, and that very high precision will be needed to tell them apart.}

To learn what the state of a black hole is, we might make measurements with many different probes over a long period of time.
This is in analogy with measuring higher moments to determine a function.
We might also ask statistical questions like
``How should a set of $K$ probes be designed to optimally classify the set of black holes of mass $M$ into $K$ classes?''.
Our discussion has focused on typical microstates of fixed energy.
It is easy to generalize to typical states in which other charges are also fixed.
For example, fixing the R-charges to take non-zero values will constrain the number of $X$ fields minus the number of $X^\dagger$ fields appearing in the operator.
For states with small R-charges, these can be regarded as introducing a small number of defects (a surplus of one of the fields) into a typical (therefore chargeless) random string in all possible ways.
Of course, we could consider ensembles of operators in which more intricate patterns are fixed ({\em e.g.}, one can consider operators in which the pattern $XYYZ$ occurs twice as frequently as the pattern $YXXZ$.)
It is possible that such restrictions, which are not related to
specifying global charges, can be associated to the specification of
quantities such as higher moments of fields in the dual gravitational
description of black hole microstates as proposed by Mathur {\em et al} \cite{Mathur,omm,mathura,mathur}.

Overall, the main lesson is that almost all probes yield almost no information about the structure of a black hole microstate.
Thus, even if there is no fundamental information loss from an underlying pure heavy state in gravity, it will be extraordinarily difficult to identify what the state is.
Indeed, the only information about the state operator that we can readily extract are conformal dimension, spin, and R-charge.
On the gravity side, these correspond to mass, angular momenta, and global charges.
In an effective picture of probes with limited energies making measurements in limited regions of space and time, the standard classical picture of black hole with a causally disconnected region should re-emerge.
It is difficult to explicitly demonstrate all of this for Schwarzschild black holes because the underlying quantum states have not been constructed in spacetime and are difficult to analyze even in the dual field theory.
Therefore, in Sections 3, 4 and 5 we will demonstrate how our considerations apply to half-BPS states of the theory which do not renormalize, and for which all computations, both in gravity and in the dual field theory, can be carried out explicitly.
We will see that almost all heavy half-BPS states are described microscopically as a quantum ``foam'', whose universal effective description at low energies is a certain singular spacetime.
The structure of the computations is identical to the discussion above which gives us some confidence that this is the correct picture even for Schwarzschild black holes.

\section{1/2-BPS States: Field theory}

Half-BPS multiplets of the $\CN =4$ Yang-Mills theory transform in $(0,p,0)$ representations of the $SO(6)$ R-symmetry group.
Highest weight states in each multiplet can be constructed as gauge-invariant polynomials in a complex scalar field $X$.
Since the superfield $X$ has conformal dimension $\Delta=1$ and carries R-charge $J=1$ under a $U(1) \subset SO(6)$ each highest-weight half-BPS state has $\Delta = J$.

A convenient method for constructing an orthonormal basis of half-BPS operators utilizes an isomorphism between representations of $U(N)$ and the symmetric group \cite{CJR,Berenstein}.
It is an old observation due to Frobenius that the representations of the symmetric group are in one-to-one correspondence with the partitions of the integers.
These are conveniently considered using Young (Ferrers) diagrams.
Consider a Young diagram $T$ with $k$ boxes encoding one such representation of the symmetric group $S_k$.  For example, for $k=9$ we might have
\begin{equation}
T = \YoungTab[-1][\tiny]{{1,2,3,4}{5,6}{7,8}{9}}
\Label{tableau1}
\end{equation}
where we label the boxes of the Young diagram with integers $1,\ldots,k$.
We can construct an associated Yang-Mills operator
\begin{equation}
\CO_T(X) = C_T \sum_{\sigma(T)} % (-1)^{a(\sigma)}
\chi_R(\sigma)
X^{i_1}_{i_{\sigma(1)}} X^{i_2}_{i_{\sigma(2)}}
\cdots
X^{i_k}_{i_{\sigma(k)}}
\Label{tabop1}
\end{equation}
where the sum runs over permutations of $\{1,\ldots, k\}$ and $\chi_R(\sigma)$ is the character (trace) of the permutation $\sigma$ in the representation $R$ of the unitary group encoded by the same diagram $T$.
$\CO_T(X)$ is a {\em Schur polynomial} in the $X$s.
Such polynomials supply a complete basis for the symmetric polynomials of degree $k$.\footnote{
Strictly speaking we are describing an independent basis of operators in a $U(N)$ theory.
The vanishing trace of $SU(N)$ matrices produces relations between some of the $\CO_T$ (see \cite{koch}).
At large-$N$ this will not change our results much.}
This is just the trace of $X$ in the representation $R$.
The coefficient $C_T$ is chosen to normalize the two-point function so that $\langle 0 | \CO_T(x) \CO_T(y) | 0 \rangle = 1 /|x -  y|^{2\Delta}$.
All half-BPS operators can be written as (descendants of) linear combinations of the $\CO_T$.

It can be shown that the constraint $\Delta = J$ implies that the
half-BPS states can also be constructed as excitations of a
Hermitian matrix in a harmonic oscillator potential
\cite{CJR,Berenstein}. This can in turn be diagonalized to a
theory of $N$ fermions $\{q_1,\ldots,q_n\}$ in a harmonic
potential. The ground state of this system, the filled Fermi sea,
consists of fermions with energies $E^g_i = (i-1)\hbar\omega +
\hbar\omega/2$ for $i=1,\ldots, N$. Every excitation above this
vacuum corresponds to a half-BPS state. Let the energies of the
fermions in an excited state be $E_i = e_i  \hbar\omega +
\hbar\omega/2$, where the $e_i$ are unique non-negative integers.
(The uniqueness is necessary because of the exclusion principle.) For the matrix model associated with the
half-BPS states of $\CN =4$ SYM we can simply choose $\omega=1$ in
view of the conformal invariance of the theory.  Each excited
state can be parametrized by a set of integers representing the
excitation energy of the $i$-th fermion over its energy in the
ground state:
\begin{equation}
r_i = \frac1\hbar ( E_i - E_i^g ) = e_i - i + 1 \,.
\Label{rowlengthdef}
\end{equation}
The $r_i$ form a non-decreasing set of integers $r_N \geq r_{N-1} \geq \cdots r_i \geq 0$, which can be encoded in a Young diagram $T$ in which the $i$-th row has length $r_i$.
For example,
\begin{equation}
\{4,3,1,1\}
\ \ \ \
  \Longrightarrow
\ \ \ \
\Young[-1]{4311}
\Label{tableau2}
\end{equation}
In our conventions the topmost row of the diagram is the longest.
It is convenient to introduce another variable $c_j$ with $1\leq j \leq N$ so that \cite{nemani}
\begin{equation}
c_N = r_1 ~~~~;~~~~ c_{N-i} = r_{i+1} - r_{i} ~~~~;~~~~ i=1,2,\ldots, (N-1) \, .
\Label{columnvars}
\end{equation}
The variable $c_j$ counts the number of columns of length $j$ in the diagram associated to $\{r_N,r_{N-1}, \ldots, r_1\}$.
A useful relation is
\begin{equation}
r_{i+1} = e_{i+1} - i = c_{N - i} + \ldots + c_N \, .
\Label{rowcolrel}
\end{equation}
The fully anti-symmetrized wavefunction for the $N$ fermion system can be expressed as a Slater determinant:
\begin{equation}
\Psi_T(\vec{q}) = \frac{1}{\sqrt{N!}} \det\left( \Psi_{e_i}(q_j)
\right)~~~~;~~~ \Psi_n(\lambda) = A(n)
H_n(\lambda/\sqrt\hbar)\,e^{-\lambda^2/2\hbar}  \, .
\Label{wavefunctions}
\end{equation}
Here $H_n$ is a Hermite polynomial, and $A(n)$ normalizes the single particle wavefunction.
Happily, the state $\Psi_T$ associated to a Young diagram $T$ turns
out to be exactly the same state as created by the operator $\CO_T$
associated to the same Young diagram via (\ref{tabop1}).
This is because the asymptotic behavior of the Schur polynomial associated to a Young diagram matches the asymptotic behavior of the Slater determinant \cite{CJR,Berenstein}.
The conformal dimension of the operators, $\Delta$, is equal to the number of boxes $k$ appearing in the diagram $T$.

Following the reasoning in Section 2, we are interested in the structure of typical half-BPS states of very large charge $\Delta = J = N^2$.
The energy in the excitation of the fermions is comparable to the energy in the Fermi sea.
In analogy with the situation for Schwarzschild black holes, we will show that almost all such half-BPS states lie close to a certain limiting Young diagram.
Because of the large amount of supersymmetry we do not expect these states to have a degeneracy large enough to be associated with a classical horizon in the dual spacetime description.
Nevertheless, half-BPS states are related to extremal charged black holes that are on the verge of producing a horizon \cite{myers}.

%%%%%%%%%%%%%% CANONICAL DISCUSSION %%%%%%%%%%%%%%%%%%%%%%%%

\subsection{Structure of typical states}

Highly excited states of the large-$N$ free fermion system can reliably be studied in a canonical ensemble in which temperature rather than energy is held fixed.
By standard reasoning the error between the canonical and microcanonical approaches will be small in the thermodynamic (large-$N$) limit in which we are principally interested.
Thus we will use the free fermion formulation to study the typical shape of Young diagrams with $\Delta$ boxes.
Related studies were carried out in \cite{buchel,nemani}.

Using the notation introduced at the beginning of Section 3, the canonical partition function of $N$ fermions in a harmonic potential is given by
\begin{equation}
\tilde{Z} = \sum e^{-\tilde{\beta} E} =
\sum_{0 \leq e_1 < e_2 \cdots < e_N} e^{-\tilde{\beta}
\hbar \omega \sum_i (e_i + 1/2) }
\Label{canpart1}
\end{equation}
with $\omega = 1$ here.
The restricted sums over the excitation energies $e_i$ (or equivalently the row-lengths $r_i= e_i -i +1$) can be replaced by unrestricted sums over the variables $c_j$ in (\ref{columnvars}) that count the number of columns of length $j$.
Carrying out this transformation of variables and setting
$Z = \tilde{Z} e^{\tilde{\beta}\hbar\omega N^2/2}$ to remove the
irrelevant vacuum energy, the partition function becomes
\begin{equation}
Z =  \sum_{c_1, c_2,\dots ,c_N=1}^\infty
e^{-\tilde\beta \hbar\omega \sum_j j c_j}    =
\prod_{j=1}^N {1 \over 1 - e^{-\beta j}},
\Label{canpart2}
\end{equation}
with $\beta = \tilde{\beta} \hbar \omega$.
Defining
\begin{equation}
q = e^{-\beta},
\Label{qdef}
\end{equation}
the temperature of the ensemble is fixed by requiring\footnote{
Note that  the temperature is simply a Lagrange
multiplier fixing the energy of our ensemble of half-BPS states.
We are not studying a thermal ensemble in the Yang-Mills theory
which would necessarily break supersymmetry.}
\begin{equation}
  \langle E \rangle = \Delta = q\, \frac{\partial}{\partial q} \log Z(q) =
  \sum_{j=1}^N \frac{j\,q^j}{1-q^j}~.
 \label{eq:meane}
\end{equation}
In the large-$N$ limit we can approximate the sum on the right hand side by an integral, provided that the contribution from the limits is not too large.  
Assuming this to be true, as we will check self-consistently later, we have
\begin{equation}
\Delta \approx \int_{j=0}^N dj \, {j q^j \over 1 - q^j}
= {\pi^2 - 6N (\log q) \log(1 - q^N) - 6\,  {\rm Li}_2(q^N) \over 6 (\log q)^2}
= \frac{\Li_2(1-q^N)}{(\log q)^2}
\end{equation}
Here ${\rm Li}_2(x) = \sum_n x^n/n^2$ is the dilogarithm function.
Since $q = e^{-\beta}$, depending on how $\Delta$ scales with $N$,
in the large-$N$ limit, $q^N$ can approach $0$, $1$, or a finite
number in between.

\paragraph{On the scaling of temperature with $N$:} We can determine
the scaling of the temperature in the ensemble with $N$ by estimating
the number of microstates of a given conformal dimension $\Delta$.
The exact number of such microstates is given by the number of
partitions of $\Delta$ into at most $N$ parts.  (Equivalently, by the
number of partitions of $\Delta$ into parts no bigger than $N$.) We
shall denote this number by $p(\Delta,\,N)$, and so the entropy will
be given by
\begin{equation}
  S = \log\,p(\Delta,\,N)~.
\end{equation}
Clearly, if $\Delta\leq N$, $p(\Delta,\,N)=p(\Delta)$ where
$p(\Delta)$ stands for the unrestricted number of partitions of
$\Delta$. When $\Delta > N$, and in the large-$N$ limit, we can
still obtain an upper bound on the scaling of the number of states with $N$
% estimate the $N$ scaling dominant dependence in the entropy
by applying the Hardy-Ramanujan formula \cite{hr}
\begin{equation}
S \approx \log\,p(\Delta)\approx 2\pi \sqrt{\Delta/6} - \log \Delta
- \log \sqrt{48} + {\cal O}(1/\sqrt{\Delta}).
\end{equation}
Using this expression we can extract the leading $N$ dependence in the
temperature by computing the variation of the entropy with respect to
the conformal dimension $\Delta$:
\begin{equation}
  \frac{1}{T} \sim \frac{1}{\sqrt{\Delta}} + \CO(\Delta^{-1})~,
\end{equation}
where we have neglected coefficients of order one, \ie those not scaling with
$N$, since in any case the right partition problem of interest is not computed by
$p(\Delta)$. 
Nevertheless this is a useful estimate because in large-$N$, $p(\Delta)$ and $p(\Delta,N)$ will differ by a factor of order unity.
Thus we learn that for typical operators of conformal
dimension $\Delta\sim N^2$, the temperature should scale linearly with
$N$. Equivalently, we are dealing with an ensemble in which $q^N$
approaches a finite number, even in the large-$N$ limit.

Given the above microcanonical estimation, we shall assume that
at large-$N$ we can write $\beta = \alpha/N$ for some constant $\alpha$.
Then
\begin{equation}
\Delta = N^2 \,  \left( \frac{{\rm Li}_2(1-e^{-\alpha})}{\alpha^2} \right) \equiv \gamma \, N^2
\Label{alphaeq}
\end{equation}
In other words,
\begin{equation}
\beta = {\alpha \over N} ~~~~~\Longrightarrow~~~~~ \Delta = \gamma N^2 \, .
\Label{betasoln}
\end{equation}
Having fixed $\beta$ in this way we can compute the shape of the typical Young diagram with $\Delta$ boxes.

The expected number of columns of length $j$ is
\begin{equation}
  \langle c_j \rangle = \frac{q^j}{1-q^j}~,
 \label{eq:meanrj}
\end{equation}
from which we derive the expected rowlength
\begin{equation}
\langle r_i \rangle = \sum_{j=0}^{i-1} \frac{q^{N-j}}{1-q^{N-j}}~.
 \label{eq:meanirow}
\end{equation}
In the large-$N$ limit overwhelmingly many half-BPS states will have associated diagrams that lie arbitrarily close to the limit shape with rows of length $\langle r_i \rangle$.
Indeed, the diagrams that do not lie along the limit shape have vanishing likelihood.
It is interesting that the columns of the Young diagram, which are interpreted as giant graviton D-branes \cite{states,CJR,holey,BHLN,Berenstein,BBFH}, are acting like a gas of bosons despite the fact that we are studying the excitations of a fermionic system.
It is well-known that small fluctuations of a Fermi sea in the $c=1$ and AdS/CFT contexts can be bosonized to described small bosonic fluctuations of spacetime \cite{llm,c=1,dhara}.
We are explaining  an interesting kind of bosonization that produces bosonic D-branes from collective fermionic excitations.

\paragraph{Entropy:} The entropy of the above canonical ensemble
can be obtained in the usual way by means of a Legendre transform
of the free energy. This is a straightforward calculation, and the
result is that for the scaling regime with $\Delta=\gamma N^2$
and $\gamma$ kept fixed as $N\rightarrow \infty$ the entropy becomes
\be S(\Delta,N)=\frac{N}{\alpha} \left( \frac{\pi^2}{6} - {\rm
Li}_2(e^{-\alpha}) \right) + \alpha \frac{\Delta}{N},
\label{canentrop}
\ee
with $\alpha$ the solution of (\ref{alphaeq}).
To get an idea of the extent to which the constraint on the number of rows reduces the number of partitions,
we take the example with $\gamma=1$, \ie $\Delta=N^2$, and compare $S(\Delta,N)$ with the entropy
$S_0(\Delta)=\log p(\Delta)$ of unconstrained partitions with $\Delta$ boxes.
For $\gamma=1$ one can numerically solve (\ref{alphaeq}) and one obtains $\alpha \simeq 0.81465$.
Inserting this into (\ref{canentrop}) yields $S\simeq 2.214 N$.
The number of unconstrained partitions is given by the Hardy-Ramanujan formula \cite{hr}
\begin{eqnarray}
p(\Delta)\approx \frac{e^{2c\,\sqrt{\Delta}}}{4\sqrt{3}\, \Delta} \left(1 - \frac{1}{2c\sqrt{\Delta}} \right);
&& c = \frac{\pi}{\sqrt{6}}.
\Label{cardy}
\end{eqnarray}
Thus $S_0(\Delta)\simeq 2\pi\sqrt{\Delta/6} \simeq 2.565 N$, which shows that fixing the number of
rows to be $N$ cuts down the entropy by a factor of about $0.86$. In
any case, this numerical analysis shows that indeed the dominant $N$
dependence in the entropy is correctly captured by the Hardy-Ramanujan
formula as  above.

As an aside, we briefly compare this to the asymptotic formula \cite{erdos}
\begin{equation} \label{erdo}
p(\Delta,N) = p(\Delta) \exp\left( -{1 \over c} e^{-cx} \right)
\end{equation}
where this is valid in the limit where $N,\Delta\rightarrow \infty$ keeping
\begin{equation}
x = {N \over \sqrt{\Delta}} - \frac{\log \Delta}{2c}
\end{equation}
fixed.\footnote{By keeping $x$ fixed here we mean that this
parameter does not scale with $N$ as the limit is taken, but is
otherwise free to run from $-\infty$ to $\infty$.}
In this regime $N$ grows faster than $\sqrt{\Delta}$, namely $N$ grows as $\sqrt{\Delta} \log
\Delta$ instead. If we nevertheless insert $\Delta=N^2$ in (\ref{erdo}) we obtain an entropy
$S\simeq 2.349 N$ which is $0.92$ times the unconstrained entropy. Either way, we find that the numbers of
constrained and unconstrained partitions are quite close to each other.

\paragraph{Temperature:}
We can also formally derive a temperature by taking the derivative of entropy with
respect to the energy $\Delta$. Since $q=\exp(-\alpha/N)$, it comes as no surprise that this yields
\begin{equation}
  \frac{1}{T} = \frac{\partial S}{\partial \Delta} = \frac{\alpha}{N}
 \label{eq:cardyt}
\end{equation}
There is no physical temperature here as we are dealing with half-BPS states.
Rather $T$ should be understood as the effective temperature of a
canonical ensemble that sums over half-BPS states with energy sharply
peaked at $\Delta$.\footnote{One can show more generally that if $\Delta$ scales as $N^\xi$, the temperature goes as $N^{\xi/2}$.}

\paragraph{Limit shape: }
Let us introduce two coordinates $x$ and $y$ along the rows and columns of the Young diagram.
In our conventions, the origin $(0,0)$ is the bottom left corner of the diagram, and $x$ increases going up while $y$ increases to the right.
In the fermion language, $x$ labels the particle number and $y$ its excitation above the vacuum.
The expression (\ref{rowcolrel}) can then be written as
\begin{equation}
   {\bf y}(x) = \sum_{i=N-x}^N\, c_i ~.
\end{equation}
Taking the expectation value,
\begin{equation}
 \langle {\bf y}(x) \rangle =
  y(x) = \sum_{i=N-x}^N\,\langle c_i \rangle~.
 \Label{limitdef}
\end{equation}
In the large-$N$ limit, we can treat $x$ and $y$ as continuum variables, and the
summation becomes an integral\footnote{
More carefully, we take $N \to \infty$ while $\hbar \to 0$ such that the Fermi level $\hbar\, N$ is kept fixed.
Rescaling $x$ and $y$ appropriately leads to the integral.}
\begin{eqnarray}
  y(x) &=& \int_{N-x}^N di\ \langle c_i\rangle 
\nonumber \\
  &=& \frac{\log (1-e^{-\beta N})}{\beta}
% + \frac{e^{-\beta N}}{1-e^{-\beta N}}
- \frac{\log (1-e^{-\beta (N-x)})}{\beta} \nonumber \\
  &\equiv & C(\beta,N) - \frac{\log (1-e^{-\beta (N-x)})}{\beta}~.
\Label{continuum}
\end{eqnarray}
After some elementary algebra, this curve can be rewritten as
\begin{equation}
  q^{N-x} + t(\beta,N) \, q^{y} =1 ~~~~;~~~~ t(\beta,N) = q^{-C(\beta,N)} \, .
 \Label{eq:canlshape}
\end{equation}

The precise meaning of the limit shape is that in the large-$N$ limit, and after a suitable rescaling of
the Young diagram, the boundary of the Young diagram approaches the limit curve with probability one.
In other words, in the large-$N$ limit almost all operators with conformal dimension $\Delta$ will have associated Young diagrams that are vanishingly small fluctuations about this curve.\footnote{
Our limit shape (\ref{eq:canlshape}) differs from the limit shapes of Young diagrams that appear in
explicit instanton calculation in ${\cal N}=2$ super Yang-Mills theory \cite{nikita} and in computations
of certain Gromov-Witten invariants \cite{okounkov}.
This is because we are averaging over diagrams using the uniform measure rather than the Plancherel measure.
We are treating all the partitions as {\em a priori} equiprobable, rather than assigning weights to partitions based on the dimension of the representation of the associated Young diagram.
This is the correct procedure in the present case because the volume of the residual orbits of the symmetric group $S_N$ that remain after gauge fixing the original $U(N)$ symmetry have been accounted for in the way we parametrize free fermion excitations.}
Limit shapes for Young diagrams have also been discussed by Vershik \cite{vershik} as part of an extensive mathematical literature on random partitions of integers.

When we approximate the sum in (\ref{continuum}) by an integral, one may worry that we miss some important  aspect of the physics. For example, it could happen that $\langle c_N \rangle$ becomes macroscopically large of order $\sqrt{\Delta}$ (say, if for instance there were some version of Bose-Einstein condensation in the theory).   For our ensembles, this only happens in certain limiting cases, which we deal with as necessary.

It is convenient to rewrite the limit curve as
\begin{equation}
  q^{-x}  \, q^{N} + q^{{y}}\,\left(1-q^N\right) = 1~.
 \Label{eq:fcanlshape}
\end{equation}

As we have seen, if $\Delta = \gamma N^2$, the inverse temperature goes as $\beta = \alpha/N$.
The above curve then becomes:
\begin{equation}
  e^{-\alpha}\,q^{-x} + (1-e^{-\alpha})\,q^{{y}} =1~.
\end{equation}
The limit shape simplifies in two natural regimes.
First, we focus on the base of the limit shape, that is, at $x\ll N$.
In this regime, the limit curve behaves like a straight line:
\begin{equation}
  {y}(x) =
  \frac{e^{-\alpha}}{1-e^{-\alpha}}\,x~, \quad \frac{x}{N}\ll 1~.
 \Label{eq:lhypcurve}
\end{equation}
Secondly, we focus on the top of the limit diagram.
Introducing the variable $\tilde{x}=N-x \ll N$, we find logarithmic behavior:
\begin{equation}
  {y}(x) = -\frac{N}{\alpha} \log
  \left(\frac{{\alpha}}{1-e^{-\alpha}}\,\frac{\tilde{x}}{N}\right)~.
 \Label{eq:loghypcurve}
\end{equation}

%%%%% DISCUSSION OF FLUCTUATIONS %%%%%%%%%%%%%%%%

\paragraph{Fluctuations: }
The accuracy of the limit shape as a description of typical states is determined by the size of fluctuations in individual states.
The variance in the number of columns of length $j$ turns out to be
\begin{equation}
{\rm Var}(c_j) = \langle ( c_j - \langle c_j \rangle)^2 \rangle = \langle c_j^2 \rangle - \langle c_j \rangle^2
= {q^j \over (1 - q^j)^2} = {\langle c_j \rangle \over 1- q^j}
\Label{colvar}
\end{equation}
just like the fluctuations of population numbers in a bosonic gas.
Thus the standard deviation divided by the mean is
\begin{equation}
{\sigma(c_j) \over \langle c_j \rangle} =
\left( {1 \over \sqrt{ \langle c_j \rangle}} \right) {1 \over \sqrt{1 - q^j}} =
\left( {1 \over \sqrt{ \langle c_j \rangle}} \right) {1 \over \sqrt{1 - e^{-\alpha j / N}}}
\end{equation}
where the last equality sets $\beta = \alpha/N$.
Thus, we are seeing the standard $\sqrt{n}$ random fluctuations in counting populations, enhanced by a factor representing the tendency of bosons to condense to low-energy states.
This factor matters most for the number of short columns (small $j$).

It is also interesting to compute the fluctuations in the limit shape (\ref{eq:fcanlshape}).
In order to calculate this using (\ref{limitdef}), we use the fact that the numbers of columns of different lengths are uncorrelated variables:
\begin{equation}
\langle c_i\,c_j \rangle = \langle c_i \rangle\langle c_j \rangle\quad i\neq j \, .
\end{equation}
Using this in the definition of the variance gives
\begin{equation}
{\rm Var}\, y(x) = \langle( {\bf y}(x) - \langle {\bf y}(x) \rangle)^2 \rangle
= \sum_{i= N -x}^N {\rm Var}(c_i) \, .
\end{equation}
Taking the continuum limit as in (\ref{continuum}) we find
\begin{equation}
{\rm Var}\, y(x) =
- {1 \over \beta} {1 \over 1 - e^{-\beta N}} +
{1 \over \beta} {1 \over 1 - e^{-\beta(N - x)}}.
\Label{varcont}
\end{equation}
We can assemble this into a probability distribution describing the likelihood of small fluctuations around the limit shape
\begin{equation}
\Pr[{\bf y}(x)]
= C  \exp\left[- \int_0^N dx {({\bf y}(x) - y(x))^2 \over 2 {\rm Var}\, y(x)} \right]
\Label{limitshapedist}
\end{equation}
The normalization constant $C$ is a functional determinant of
${\rm Var}\, y(x)$ in the usual way and ensures that
$\int \CD y(x) \,  \Pr[{\bf y}(x)] = 1$.   After a suitable conversion
of the variable ${\bf y}(x)$ into a probability distribution, this
expression could be compared to Sanov's theorem (\ref{sanov}) which
described the probability of having a large deviation for a particular
Schwarzschild microstate from the typical state.

\subsection{Error estimates}

It is useful to estimate the error incurred in our considerations by replacing sums by integrals and by taking a canonical rather than a microcanonical approach.
Different ensembles typically agree at leading order in an expansion in $1/E$, but the subleading terms will typically be different.
To get an idea of the magnitude of these errors, we turn to the simple case of diagrams with $\Delta$ boxes and no constraints on the number of columns.
This is reasonable to study because, as we showed in Section 3.1, the restriction on the number of rows makes only a minor difference to the overall degeneracy of states.

Taking the logarithm of the Hardy-Ramanujan formula \eref{cardy}, we see that
\be \label{ne1}
\log p(\Delta) = 2\pi \sqrt{\Delta/6} - \log \Delta - \log \sqrt{48} + {\cal O}(1/\sqrt{\Delta}).
\ee
When we replace sums by integrals, we recover only the first term in this expansion, and from this we see that the typical error introduced by replacing sums by integrals goes as $\log E/ \sqrt{E}$. Below we will see
another example where the difference between the sum and integral yields an error of order $1/\sqrt{E}$.
Though we have not extensively studied the errors in all ensembles discussed in this paper, it seems
very likely that similar estimates apply there as well.

In a canonical analysis, one computes the partition function $Z(q)=\sum_\Delta p(\Delta) q^\Delta$:
\be \label{ne2}
Z=\sum_\Delta \exp \left(\Delta \log q +2\pi \sqrt{\Delta/6} - \log \Delta - \log \sqrt{48} + {\cal O}(1/\sqrt{\Delta}) \right).
\ee
This partition function peaks at $\Delta=\Delta_0$, where
\be \label{ne3}
\log q = -\frac{\pi}{\sqrt{6\Delta_0}} + \frac{1}{\Delta_0} + {\cal O}(\Delta_0^{-3/2}).
\ee
The variance is obtained by differentiating $\log Z$ twice at $\Delta=\Delta_0$ and then taking the inverse:
\be
\sigma^2 = -\left( \left. \frac{\partial^2 \log Z}{\partial \Delta^2} \right|_{\Delta=\Delta_0} \right)^{-1}.
\ee
This works out to be
\be \label{ne4}
\sigma^2 = \frac{\sqrt{24 \Delta_0^3}}{\pi} + \frac{24 \Delta_0}{\pi^2} +{\cal O}(\sqrt{\Delta_0}).
\ee
As a check, one can compute $\sigma^2$ simply via the standard expression for the variance of the energy in a canonical ensemble
\be
\sigma^2 = \langle ( \Delta - \langle \Delta \rangle )^2 \rangle =
\sum_{n>0} \frac{n^2 q^n}{(1-q^n)^2} \sim \frac{2\zeta(2)}{(\log q)^3}
\ee
which reproduces the leading term in (\ref{ne4}) when we use (\ref{ne3}).
From the above we infer that in this case the mistakes that are made when replacing sums by integrals are all of order $1/\sqrt{\Delta}$.

A slightly more complicated example is the expectation value of the number of columns with $j$ boxes.
The canonical ensemble answer is very simple,
\be \label{ne5}
\langle c_j \rangle = \frac{q^j}{1-q^j} .
\ee
The microcanonical answer is on the other hand given by\footnote{
A given partition with precisely $m$ columns of length $j$ will appear a total of $m$ times in the sum evaluated in the numerator, as required.}
\be
\label{ne6} c_j = \frac{\sum_{k>0} p(\Delta-kj)}{p(\Delta)} .
\ee
To evaluate this, we use the asymptotic form of $p(\Delta)$ in (\ref{ne1}), and expand the exponent in $kj$.
We discover that
\be
\sum_{k>0} p(\Delta-kj) \sim
\sum_{k>0} \exp\left( -\pi \frac{kj}{\sqrt{6\Delta}} +\frac{kj}{\Delta} +\ldots \right),
\ee
where the dots indicate terms of higher order in $kj$ or terms suppressed by higher powers of $\Delta$.
If we keep only the leading term, we can do the sum exactly and recover (\ref{ne5}) with $q=\exp(-\pi/\sqrt{6\Delta})$, the leading answer to (\ref{ne3}).
The two obvious sources of mistakes are dropping other terms linear in $kj$, which will shift the value of $q$ to a slightly different value, and dropping the quadratic and higher terms in $kj$.
It seems that the main source of error arises from the interference of quadratic terms with linear ones.
But these barely have any impact unless $j$ starts to become of order $\Delta^{3/4}$, so that for $k=1$ there is some suppression from the quadratic terms.
Thus we reach the preliminary conclusion that the microcanonical answer and the canonical answer differ by a factor of the form
\be
\left(1 + {\cal O}(\frac{j}{\Delta^{3/4}},\frac{1}{\sqrt{\Delta}}) \right).
\ee
We have checked numerically that this is the right behavior.

The next step is to consider the variance in the distribution of the number of columns of a given length.
In Section 3.2 we showed that this was
\be
\sigma^2 =
\frac{q^j}{(1-q^j)^2}. \label{ns1}
\ee
The exact microcanonical answer is\footnote{
A given partition with $m$ columns of length $j$ will appear in the sum in the numerator once for each $1 \leq k \leq m$.
Thus it will contribute a weight $\sum_{k=1}^m (2k-1) = m^2$, exactly as required.}
\be
\langle c_j^2 \rangle  =  \frac{\sum_{k>0} (2k-1) p(\Delta-kj)}{p(\Delta)} .
\ee
With the same manipulations as above we find that this equals
\be
\frac{q^j(1+q^j)}{(1-q^j)^2}  \left(1 + {\cal O}(\frac{j}{\Delta^{3/4}},\frac{1}{\sqrt{\Delta}}) \right)
\ee
so finally we find that the microcanonical answer for $\sigma^2$ is given by
\be \sigma^2 = \frac{q^j}{(1-q^j)^2} \left(1 + {\cal O}(\frac{j}{\Delta^{3/4}},\frac{1}{\sqrt{\Delta}}) \right) .
\ee
The corrections are found in the same way as above.
The spread in the expectation value of $c_j$ is therefore
\be
\frac{\sigma}{\langle c_j \rangle} = q^{-j/2}\left(1 + {\cal O}(\frac{j}{\Delta^{3/4}},\frac{1}{\sqrt{\Delta}}) \right) .
\ee
This illustrates how the number of columns of each length is distributed as we probe all the states in a microcanonical ensemble.

In the subsequent considerations in this paper we will ignore corrections such as the ones studied in this section, but it is important to keep in mind that they are there.

\subsection{The superstar ensemble}

In this section, we analyze an ensemble of Young diagrams in which not
only the number of boxes $\Delta$ and the number of rows $N$ are held fixed,
but also the number of columns $N_C$.  Physically, the last constraint amounts to holding fixed the number of D-branes in spacetime \cite{states, CJR}.\footnote{At a more technical level, introducing the
parameter $N_C$ allows us to have some control over the way in
which this number scales with $N$.}    One might expect that this ensemble includes a
microscopic description of the extremal ``superstar'' configuration
\cite{myers}, since this spacetime involves a constraint relating the total energy, the amount of flux and the number of D-branes.  We will indeed find a  description of the superstar  once we restrict the three  parameters to obey $\Delta=N N_C/2$. For arbitrary values of the
three parameters we find a finite ``temperature'' deformation of
the superstar.

The addition of the column number constraint changes the statistical properties
of the system considerably.  In the previous ensemble,
the average energy could grow without bound by increasing
the temperature of the ensemble. The larger the
energy, the larger the entropy of the system. In the present
set-up, this is no longer true. Given an ensemble characterized by
the pair $(N,\,N_C)$, the conformal dimension $\Delta$ lies in a
finite interval :
\begin{equation}
  \langle E \rangle = \Delta \,\in \, [N_C,\,N\cdot N_C]~.
\end{equation}
The lower bound corresponds to a one row Young diagram with
$N_C$ boxes; whereas the upper bound corresponds to a rectangular
diagram in which all $N$ rows have $N_C$ boxes. Clearly, both bounds
have a unique microstate, and so we can conclude our system has
vanishing entropy in both situations.

Intuitively, one expects that as one increases the energy slightly
above $N_C$, the entropy increases, heating up the system to a finite
positive temperature. This behavior should continue until the entropy
reaches a maximum, which is achieved at $\Delta_{{\rm s}}=N_C(N+1)/2$, when
half of the allowed boxes are filled in.
If the energy goes beyond this
value, we expect the entropy to start decreasing as the degeneracy of
microstates will decrease. Indeed, there is a one-to-one map between Young diagrams with
$\Delta$ boxes and $N_C(N+1)-\Delta$ boxes, and therefore $S(\Delta)=S(N_C(N+1)-\Delta)$.
This shows that in this regime our system
will have {\it negative} temperature.\footnote{We recall that the
temperature in these ensembles is not physical. It is merely a
parameter to characterize typical states, which in this case allows us
to correctly describe the situation in which an increase in energy
causes the entropy to decrease.} As the energy continues to grow, the
absolute value of the temperature will decrease, reaching a zero value
for the upper bound. This discussion strongly suggests that the system
achieves an infinite temperature when the entropy is maximized.  Below we will see that this is precisely what happens, and in Section~4 we will show that the maximum entropy configuration is described in gravity as the superstar of \cite{myers}.

The description of typical states will be based on the canonical ensemble analysis in Section 3.2, with an additional Lagrange multiplier to implement the  constraint on the number of columns in the Young diagram.  Using the same notation as before, the partition function for the system, once we have removed the energy contribution from the vacuum, is:
\begin{equation}
  Z = \sum_{c_1,c_2,\cdots,c_N =1}^\infty e^{-\beta\sum_j jc_j - \lambda\left(\sum_j c_j - N_C\right)}~,
\end{equation}
where $\lambda$ stands for the Lagrange multiplier.
(Recall that $c_j$ counts the number of columns of length $j$.)
Thus we will require the condition
\begin{equation}
  \frac{\partial Z}{\partial\lambda} = 0\,\, \Leftrightarrow \,\, \sum_j c_j = N_C~.
\end{equation}
Proceeding as before, this partition function can be written as
\begin{equation}
  Z = \zeta^{-N_C}\,\prod_j \frac{1}{1-\zeta\,q^j}~, \quad \zeta =
  e^{-\lambda}~.
\end{equation}

The temperature of the ensemble is fixed by requiring
\begin{equation}
  \langle E \rangle = \Delta = q\partial_q \log Z(\zeta,\,q) =
  \sum_{j=1}^N \frac{j\,\zeta\,q^j}{1-\zeta\,q^j}~,
 \Label{esuper}
\end{equation}
whereas the Lagrange multiplier $(\lambda)$, or equivalently $\zeta$,
will be fixed by
\begin{equation}
  N_C = \sum_{j=1}^N \langle c_j\rangle = \sum_{j=1}^N
  \frac{\zeta\,q^j}{1-\zeta\,q^j}~,
 \Label{ncsuper}
\end{equation}
where we already computed the expected number of columns of length
$j$, \ie $\langle c_j\rangle$.

We shall examine the different natural temperature regimes in this
ensemble, to check the claims made in our preliminary discussion.
For the time being, we will keep $N$ fixed and vary $\beta$.    We can also take scaling limits involving both $\beta$ and $N$.   These two ways of taking limits are not identical.   They lead to similar qualitative
conclusions but the quantitative details are different.

So for now we keep $N$ fixed and consider the {\it large}-$\beta$ expansion. If the temperature
is {\it positive}, this corresponds to the approximation
$q=e^{-\beta}\ll 1$. In such regime, both sums in (\ref{esuper}) and
(\ref{ncsuper}) are dominated by their first term. This allows us to
conclude
\begin{equation}
  \Delta \sim N_C \sim \langle c_1\rangle =\frac{\zeta q}{1-\zeta q}~,
\end{equation}
which indeed corresponds to a diagram with a single row of $N_C$ boxes
since the dominant contribution comes from columns of length one.

If the temperature is {\it negative}, corresponding to the
approximation $q=e^{-\beta}\gg 1$, the number of columns
\begin{equation}
  N_C \sim \langle c_N \rangle = \frac{\zeta q^N}{1-\zeta q^N}~,
\end{equation}
is dominated by the number of columns of length $N$. This is so
because the function $x/(1-x)$ is monotonically increasing in
$x$. Thus the corresponding diagram is a rectangular one, as
expected. To further check this interpretation, we can compute the
dominant contribution to the energy (\ref{esuper})
\begin{eqnarray}
  \Delta &=& \sum_{s=1}^\infty \zeta^s\,\sum_{j=1}^N j\,q^{sj}
\nonumber \\
  & =& \sum_{s=1}^\infty \zeta^s\left\{q^s\,\frac{1-q^{sN}}{1-q^s} +
  q^{2s}\,\frac{1-N\,q^{(N-1)s}+(N-1)q^{Ns}}{(1-q^s)^2}\right\} \nonumber \\
  &\sim & N\,\sum_{s=1}^\infty (\zeta\,q^N)^s = N\,N_C~,
\end{eqnarray}
which indeed agrees with the energy of a rectangular diagram with $N$
rows of $N_C$ boxes each.

Let us analyze the more interesting {\it high} temperature regime, or
equivalently, the small-$\beta$ expansion. We shall treat both the
positive and negative temperatures together including a sign in the
definition of the expansion parameter $(\beta)$. We are interested in
computing the entropy to second order in $\beta$, to provide evidence
for the existence of a maximum at $\beta=0$. Using the identity,
\begin{equation}
  S = \beta\,\Delta + \log Z = \beta\,\Delta - N_C\log\zeta -
  \sum_{j=1}^N \log (1-\zeta\,q^j)~,
\end{equation}
we realize that we need to work out the expansions for $\zeta$ and $\Delta$ at
second and first orders, respectively. Expanding (\ref{ncsuper}) and inverting the
corresponding equation, we can find the expression for $\zeta$:
\begin{equation}
  \zeta (N,\,N_C) = \frac{\omega}{1+\omega}\left(1 + \beta\,\frac{A_1}{N} +
  \beta^2\frac{N+1}{12} ( (N+2)-\omega(N-1)) \right)
 + \CO(\beta^3)
  \,~.
\end{equation}
In the above expression we have introduced the notation:
\begin{equation}
  \omega \equiv \frac{N_C}{N}~, \quad A_1 = \sum_{j=1}^N j =
  \frac{N(N+1)}{2}~.
\end{equation}
Expanding (\ref{esuper}), the conformal dimension at first order is
\begin{equation}
  \Delta = \omega\,A_1 - \beta\,\frac{A_1}{6}\omega(1+\omega)\,(N-1) +
  \CO(\beta^2)~.
\end{equation}
Notice that at large-$N$, the dominant energy is the one computed for
the superstar in \cite{myers}, suggesting that in the $T \to \infty$ limit the large-$N$ typical state is described in spacetime as the superstar.  We will show this explicitly in Section~4.3.  The typical states in this infinite temperature ensemble correspond to Young diagrams that are nearly triangular. That is, on average,  there is a constant gap $(\omega)$ of energy between the
excitations of the $(j+1)$-st and $j$-th fermions, the first fermion
having an average energy $\omega$ itself. Moreover, having
computed the linear $\beta$ dependence allows us to confirm that for
positive temperatures, the conformal dimension is smaller than the
superstar energy, whereas it is for negative temperatures that the
conformal dimension can exceed the latter, in agreement with our
general entropic arguments.

Finally, if the superstar ensemble maximizes the
entropy, the entropy expansion in $\beta$ should have no linear
dependence in it and its second order coefficient must be negative for
all values of $N,\,N_C$. Carrying out the computation, we obtain
 \begin{equation} \label{auxx}
  S = -\log \left( \frac{\omega^{N_C}}{(1+\omega)^{N+N_C}} \right) -\beta^2
\frac{\omega(1+\omega)}{24} N(N^2-1)
+ \CO(\beta^4)~.
\end{equation}
Since $N\geq 1$, the coefficient of $\beta^2$ is negative, for any
value of $(N,\,N_C)$, as expected. According to our discussion, we
should identify the first term above
as a microscopic derivation for the entropy of the superstar.

\paragraph{Limit shape: } Proceeding as in the previous section, in
the large-$N$ limit, we can deduce the continuum limit curve:
\begin{eqnarray}
  y(x) &=& \sum_{j=N-x}^N \langle c_j\rangle = \int_{N-x}^N \langle c_j\rangle dj\  \nonumber \\
  &=&  \frac{1}{\beta}\log \left( 1-\zeta\,q^N\right) - \frac{1}{\beta}\log\left(1-\zeta\,q^{N-x}\right) \nonumber \\
  &\equiv & C(\beta,\,\zeta) -\frac{1}{\beta} \log\left(1-\zeta\,q^{N-x}\right)~,
\Label{superlshape}
\end{eqnarray}
where our conventions for both $\{x,\,y\}$ along the Young diagram and its
origin are as before. This limit shape can be written more symmetrically as
\begin{equation}
  \zeta\,q^{N-x} + t\,q^{y-N_C} = 1~, \quad t\equiv q^{N_C- C(\beta,\,\zeta)}~.
 \Label{fsuperlshape}
\end{equation}
Since we are working in the limit of large $N,N_C$ the slight asymmetry between
$N$ and $N_C$ that we had before is irrelevant. In particular, the points $(x,y)=(0,0)$
and $(x,y)=(N,N_C)$  lie on the limit curve. Therefore
\be
t= \frac{q^N-1}{q^N-q^{-N_C}}, \qquad \zeta=\frac{1-q^{-N_C}}{q^N-q^{-N_C}}.
\ee
This is equivalent to the constraint defining the ensemble
(\ref{ncsuper}) as one may easily verify. Exactly the same limit curve also appears in
theorem~4.7 in \cite{vershik}, where it is shown that this is indeed the limit curve
if we take $\Delta\rightarrow\infty$ keeping $N/\sqrt{\Delta}$ and $N_C/\sqrt{\Delta}$ fixed.

To make the symmetry between $N$ and $N_C$ more manifest we rewrite the limit curve (\ref{fsuperlshape})
as
\be
\label{eq101}
\alpha q^{N-x} + \beta q^y = 1,
\ee
with
\be \label{def:ab}
\alpha=\zeta=\frac{1-q^{N_C}}{1-q^{N+N_C}}, \qquad
\beta =\frac{1-q^{N}}{1-q^{N+N_C}}.
\ee
The energy, or in other words $\Delta$, can be computed
from (\ref{esuper}) which is of course the same as computing
the area under the limit curve. We obtain
\be
\label{eq204}
\Delta=\frac{ {\rm Li}_2(\alpha) + {\rm Li}_2(\beta) + \log\alpha\log\beta - \pi
^2/6 }{(\log q)^2}
\ee
which is indeed symmetric under $N\leftrightarrow N_C$.
The parameter $q$ in the grand canonical partition function is nothing but $e^{-1/T}$ and
we saw above that as $T\rightarrow \infty$ the entropy reaches a maximum and is dominated by
triangular Young diagrams. It is easy to confirm this from the limit shape:
expanding eq.\ (\ref{eq101}) around $q=1$, we find that the limit shape becomes
\be
\frac{N y - N_C x}{N+N_C} (q-1) + {\cal O}((q-1)^2)=0.
\ee
To leading order this is indeed a straight line
\be
y = \frac{N_C}{N} x = \omega x
\Label{eq:triangle} \label{sline}
\ee
whose slope is ratio of the number of columns to the number of rows, similarly to what we saw above.

The Young diagram that is encoded by
the limit shape \eref{eq:triangle} admits two natural
interpretations. We can read the diagram from left to right or
from top to bottom. If we read from left to right, the columns
represent giant gravitons, D3-branes that wrap $S^3\subset S^5$.
The maximum angular momentum of any of these giant gravitons is
$N_R\le N$ but their number is unconstrained. The geometry given
by this limit shape arises as a bound state of $N_C$ giant
gravitons. The angular momentum of the $y$-th giant graviton is
$N_R - y N_R/N_C$. If we read from top to bottom, the rows
represent dual giant gravitons, D3-branes that wrap $S^3\subset
\AdS{5}$. We can have at most $N_R\le N$ of these because dual
giant gravitons are stabilized by flux, and the flux through the
last dual giant represented by the bottom row of the Young diagram
is $N-N_R$. The same geometry will arise as a bound state of the
$N_R$ dual giants. The angular momentum of the $x$-th dual giant
graviton is $y = x N_C/N_R$.\footnote{That there are two such
interpretations of the same background is a manifestation of the
particle/hole duality of the quantum Hall description of the half-BPS states \cite{beren,shahin}.  In fact this exchange symmetry even manifests itself away from the BPS limit.  It is shown in \cite{correspondence} that the entropy of the non-extremal superstars with charge $G$ in a background with $N$ units of flux is equal to the entropy of charge $N$ superstars in a space with $G$ units of flux.  This is a non-extremal incarnation of the exchange symmetry between rows and columns in the ensembles we are considering.}
As we shall see
explicitly in the next section, this geometry is the superstar. We defer further
discussion of this for the moment in order to further explore the
thermodynamic properties of this limit shape.

We briefly return to the entropy of the system. The scaling limit relevant for the limit curve
is one where $\Delta\rightarrow\infty$ while keeping $\lambda=\sqrt{\Delta}/N_C$ and $\mu=\sqrt{\Delta}/N$ fixed.
In this regime $q$ will behave as $\exp(\xi/\sqrt{\Delta})$ for some fixed $\xi$, and $\alpha$ and $\beta$ as defined
in (\ref{def:ab}) will be finite as well. Given $\lambda,\mu$,
the three functions $\xi,\alpha,\beta$ are the solutions of the three equations
\bea \alpha & = & \frac{1-e^{\lambda\xi}}{1-e^{(\lambda+\mu)\xi}} \nonumber \\
\beta & = &\frac{1-e^{\mu\xi}}{1-e^{(\lambda+\mu)\xi}} \nonumber \\
\xi^2 & = & {\rm Li}_2(\alpha) + {\rm Li}_2(\beta) + \log\alpha\log\beta - \pi^2/6. \label{eq210}
\eea
and the entropy that we obtain by Legendre transforming the free energy is
\be\label{eq208}
S(\lambda,\mu,\Delta) \simeq (-2\xi - \lambda\log\alpha - \mu\log\beta) \sqrt{\Delta}
\ee
These equations are difficult to solve in general.
However, the linear limit shape is easy to get by taking $\xi\rightarrow 0$,
and we see that $\alpha=\lambda/(\lambda+\mu)$ and $\beta=\mu/(\lambda+\mu)$.
For the linear limit shape  with $N_C$ columns and $N$ rows we therefore find that
\be
\Delta=\frac{1}{2} N N_C, \qquad
\lambda=\sqrt{\frac{2N_C}{N}}, \qquad
\mu = \sqrt{\frac{2N}{N_C}},
\ee
and the entropy is given by
\be
S\simeq \log\left[\frac{ (\lambda+\mu)^{\lambda+\mu} }{\lambda^{\lambda} \mu^{\mu}} \right] \sqrt{\Delta}
\ee
which agrees with the leading term in (\ref{auxx}).

%%%%%%%%%%%%%%%%%% START OF GRAVITY DESCRIPTION %%%%%%%%%

\section{1/2-BPS States: Effective description in gravity}

In \cite{llm} the most general type IIB geometry invariant under $SO(4)\times SO(4)\times \BR$ preserving one-half of the supersymmetries, consistent with $\Delta=J$, is determined.
We develop the correspondence between these geometries and  effective semiclassical description of half-BPS states in the dual Yang-Mills theory, particularly those spacetime(s) that describe states with $\Delta \simeq N^2$.
We argue that almost all such states have as an underlying structure a ``quantum foam'' whose universal effective description in supergravity is a certain singular spacetime that we shall dub the ``hyperstar''.
The singularity arises because the classical description integrates out the microscopic details of the quantum mechanical wavefunction.

In \cite{llm} all the relevant half-BPS supergravity backgrounds are constructed in terms of metric and self-dual five-form field strength $F_{(5)}= F \wedge d\Omega_3 + \tilde{F}\wedge d\tilde{\Omega}_3$, where $d\Omega_3$ and $d\tilde{\Omega}_3$ are the volume forms of the two three-spheres where the two $SO(4)$s are realized.
The metric is:
\bea
  ds^2 &=& - h^{-2}
  (dt + V_i dx^i)^2 + h^2 (d\eta^2 + dx^idx^i) + \eta\, e^{G} d\Omega_3^2
  + \eta\, e^{ - G} d \tilde \Omega_3^2 \label{solmetric} \\
  h^{-2} &=& 2 \eta \cosh G , \label{solmetric2} \\
  \eta \partial_\eta V_i &=& \epsilon_{ij} \partial_j z,\qquad
  \eta (\partial_i V_j-\partial_j V_i) = \epsilon_{ij} \partial_\eta z
  \label{solmetric3} \\
  z &=&{ 1 \over 2} \tanh G   \label{solmetric4}
 %  \\
%  F &=& dB_t \wedge (dt + V) + B_t dV + d \hat B ~,~~~~~~ \nonumber
%  \\
%  \tilde F &=& d\tilde B_t \wedge (dt + V) + \tilde B_t dV + d \hat { \tilde B}
% \label{4dgf2} \\
%  B_t &=& - {1\over 4} \eta^2 e^{2 G} ,~~~~~~~~~~~~~~~~~~~~~~~~~~
%  \tilde B_t = - { 1 \over 4}  \eta^2 e^{- 2 G} \\
%  d \hat B &=&  - { 1 \over 4} \eta^3 *_3 d ( { z + \half \over \eta^2 }) ~,~~~~~~~~~~~~~~~~
%  d \hat {\tilde B} = - { 1 \over 4} \eta^3 *_3 d ( { z - \half \over \eta^2})
  ~~ \label{4dgf}
\eea
where $i=1,2$ and $*_3$ is the flat space epsilon symbol in the directions $\eta,x_1,x_2$.
In addition there is a self-dual flux that depends on the function $z$.
The entire solution is evidently determined by the single function $z$, which obeys the linear differential equation
\be
 \partial_i \partial_i z + \eta \partial_\eta \left( { \partial_\eta z \over \eta} \right) =0~.
\label{eq:zequation}
\ee
The function $\Phi(\eta;\,x^1,\,x^2) = z\,\eta^{-2}$ satisfies the Laplace equation for an electrostatic potential in six dimensions that is spherically symmetric in four of these directions.
The coordinates $x_1,x_2$ parametrize an $\BR^2$, while $\eta$ is the radial coordinate in the $\BR^4$ transverse to this auxiliary six-dimensional manifold.
Note that $x_1$ and $x_2$ have dimensions of $({\rm length})^2$.
The analogy with electrostatics allowed \cite{llm} to solve for the bulk geometry in terms of the boundary condition $z(0,\,x_1',\,x_2')$ at the origin $(\eta=0)$:
\begin{equation}
  z(\eta;\,x_1,\,x_2) = \frac{\eta^2}{\pi}\int dx_1'\,dx_2'\,
  \frac{z(0;\,x_1',\,x_2')}{[(x-x')^2 + \eta^2]^2}~.
 \label{eq:zsol}
\end{equation}
$z(0;\,x_1,\,x_2)=\pm 1/2$ are the only boundary conditions compatible
with {\it non-singular} geometries. In terms of the asymptotic 
$\ads{5} \times S^5$ geometry, at points on the $(x_1,x_2)$ plane where $z=1/2 \, (-1/2)$, an $S^3$ in the $S^5$ ($\ads{5}$) shrinks to zero size.
This topologically complex two-plane is spliced smoothly into the full bulk spacetime.
Of course, other boundary conditions are also compatible with a half-BPS condition; these will give rise to singular spacetime geometries.
Notably, $|z(0;x_1,x_2)| > 1/2$ leads to closed timelike curves while $|z(0;x_1,x_2)| < 1/2$ gives null singularities \cite{caldarelli,milanesi}.

Following \cite{llm}, in order to match these solutions with states in the field theory, consider geometries for which the regions in the $(x_1,x_2)$ plane where $z = -1/2$ are compact.
We will call these regions collectively the droplet $\CD$.
The quantization of flux in the geometry leads, even in the semiclassical limit, to an identification between $\hbar$ in the dual field theory and the ten-dimensional Planck length:
\begin{equation}
\hbar ~ \leftrightarrow ~ 2\pi \ell_p^4.
\Label{hbar}
\end{equation}
(The units in this identification are unusual because of the dimensions of $(x_1,x_2)$.)
Furthermore, the area of the droplet $\CD$ is quantized and must equal $N$, which is equivalently the total amount of five-form flux, the rank of the dual gauge group, and the number of fermions in the free Fermi picture of half-BPS states:
\begin{equation}
N = \int_{\CD} \frac{d^2x}{2\pi\hbar},
\Label{Ndef1}
\end{equation}
where we used (\ref{hbar}).
The conformal dimension $\Delta$ of a given configuration (state) is computed by
\begin{equation}
  \Delta = \int_{\CD}
  \frac{d^2x}{2\pi\hbar}\,\frac{1}{2}\frac{x_1^2+x_2^2}{\hbar} -
  \frac{1}{2}\left(\int_{\CD} \frac{d^2x}{2\pi\hbar}\right)^2~.
 \Label{eq:bps}
\end{equation}
These equations have a remarkable interpretation in terms of the hydrodynamic limit of the phase space of the fermionic system used to construct half-BPS states in Section 3 \cite{CJR,Berenstein}.
Specifically, if we identify the $(x_1,x_2)$ plane with the phase plane of a single fermion in a harmonic potential,
\begin{equation}
(x_1,x_2) ~ \leftrightarrow ~ (q,p) \, ,
\end{equation}
then in the hydrodynamic limit for an $N$-particle fermionic system (\ref{Ndef1}) and (\ref{eq:bps}) precisely compute the number of fermions and the total energy over the vacuum.
In fact, the first term in (\ref{eq:bps}) calculates the total energy of a fermion droplet in phase space while the second term in (\ref{eq:bps}) subtracts off the vacuum energy.
This identification gains support because the droplet corresponding to the filled Fermi sea (\ie a circular disk) reproduces the geometry of the $\ads{5} \times S^5$ vacuum.
Based on such observations \cite{llm} proposed that the $(x_1,x_2)$
plane of their half-BPS geometries should be identified with the
single particle phase space plane of the dual fermionic system.

For our purposes it is convenient to introduce the function \cite{mandal}
\begin{equation}
  u(0;\,x_1,\,x_2) = \frac{1}{2} - z(0;\,x_1,\,x_2)~,
\end{equation}
which takes values one or zero for non-singular geometries.
For these, the energy integral (\ref{eq:bps}) may now be written as
\begin{eqnarray}
  \Delta &=& \int_{\BR^2}
  \frac{d^2x}{2\pi\hbar}\,\frac{1}{2}\frac{x_1^2+x_2^2}{\hbar}\,
  u(0;\,x_1,\,x_2)-\frac{1}{2}\left(\int_{\BR^2}
  \frac{d^2x}{2\pi\hbar}\, u(0;\,x_1,\,x_2)\right)^2~,   \Label{uDelta} \\
  N &=&   \int_{\BR^2}
  \frac{d^2x}{2\pi\hbar}\, u(0;\,x_1,\,x_2) \Label{uN} \, .
\end{eqnarray}
These expressions resemble {\it expectation values}, suggesting that the function $u$ should be identified with the semiclassical limit of the quantum single-particle phase space distributions of the dual fermions.
(Also see \cite{mandal,jevicki,japanese} which appeared while this paper was being prepared.)
We will see that $u$ does not always assume the values $0,1$; the generic spacetime description of half-BPS states will be singular.

\subsection{Proposal}
Below we shall carry out a detailed study of the semiclassical limit of the phase distributions of the half-BPS fermionic system and explain how the function $u$ arises as a coarse-grained effective description of underlying quantum dynamics.
However, even before doing this analysis, it is possible to propose a detailed map between specific Young diagrams representing half-BPS states and the corresponding semiclassical geometries, which will all be circularly symmetric in the $x_1,x_2$ plane.
Consider states whose descriptions in terms of Young diagrams have a well-defined semiclassical limit.
For the moment we will take this to mean that as $\hbar \to 0$ with $N\hbar$ fixed, the perimeter of the Young diagram approaches a smooth curve $y(x)$ with $x$ parametrizing the rows and $y$ the columns. 
Examples of such curves were derived in the previous section as limiting shapes of typical partitions in various thermodynamic ensembles.  We propose that in such situations the integral formulae (\ref{uDelta}), (\ref{uN}) extend to differential relations:
\begin{eqnarray}
  \frac{u(0;r^2)}{2\hbar}\,dr^2 &=& dx~
 \Label{eq:fmatch} \, , \\
  \frac{r^2\,u(0;r^2)}{4\hbar^2} dr^2 &=& (y(x)+x)\,dx~.
 \Label{eq:egain}
\end{eqnarray}
Here we have written $u(0;x_1, x_2)$ as $u(0;r^2)$ in terms of the radial coordinate $r$ in the $(x_1,x_2)$ plane in view of the $U(1)$ symmetry present for single Young diagram states.
In terms of the phase space interpretation of (\ref{uDelta}), (\ref{uN}), the first equation simply relates the number of particles in phase space within a band between $r$ and $r + dr$ to the number of particles as determined by the rows of the associated Young diagram.
The second equation matches the energy of the particles in phase space within a ring of width $dr$ to the energy in terms of the Young diagram coordinates.\footnote{
Recall that $x$th fermion has an excitation energy of $\hbar y(x)$ over its ground state energy $\hbar x$.
Note that $\int_0^N dx\, \hbar x = \hbar \frac{N^2}{2}$, which is indeed the total vacuum energy of $N$ fermions in a harmonic potential.
Equivalently, $x\, dx$ is the continuum version of the vacuum energy stored in fermion $x$.}

Combining equations (\ref{eq:fmatch}) and (\ref{eq:egain}), we find that
\begin{equation}
y(x) + x = r^2/(2\hbar) \, .
\Label{curveenergy}
\end{equation}
This is just a check on the consistency of our proposal, since indeed $y(x) + x$ is the total energy at a given point $x$ in the diagram, and so must match the contribution to the energy $r^2/(2\hbar)$ at a given radial distance $r$ in the $(x_1,x_2)$ plane.
Taking derivatives with respect to $x$ and using (\ref{eq:fmatch}), we derive the relation
\begin{equation}
  u(0;r^2) = \frac{1}{1 + y'} \quad
  \Leftrightarrow \quad
  z(0;r^2) = \frac{1}{2}\,\frac{y'-1}{y'+1}~.
 \Label{eq:cmatch}
\end{equation}
The right hand sides of the equations (\ref{eq:cmatch}) should be
understood to be functions of $r^2$ obtained by substituting the
known boundary curve $y(x)$ into (\ref{curveenergy}) and then
inverting to find $x$ as a function of $r^2$. We thus establish a
dictionary between the functions $u(0;x_1,x_2)$ that completely
determine classical half-BPS solutions and the slope
($y'(x)=dy/dx$) of the Young diagram of the corresponding field
theory state. The only ingredients we have used in making this
identification are: (a) a differential relation between the number
of flux quanta in gravity and the number of fermions in the dual
field theory, and (b) a differential relation between energies on
the two sides. Although it is suggestive that $(x_1, x_2)$ look
like phase space variables and in particular $r^2/2$ looks like
the Hamiltonian, we did not use this in any significant way.
Remarkably, in Section 4.2 we will show that in the semiclassical
limit, at scales large compared to $\hbar$, the exact quantum
mechanical phase space distribution for the half-BPS fermion
system precisely reproduces (\ref{eq:cmatch}). The distance from
the origin is an energy.

Despite the large amount of evidence we will present for our
proposal, one could imagine going further and trying to prove it
directly in the context of AdS/CFT correspondence. That the
$(x_1,x_2)$ plane can be identified with the phase space of a
fermionic system was already demonstrated in \cite{mandal,japanese,grant}.
From this one could go on and try to show that the energy-momentum
tensor of the dual matrix model, which couples to the bulk metric,
indeed effectively integrates out all but one of the fermions, and
one could perhaps show that higher, massive string modes are
necessary in order to reconstruct the detailed structure of an
arbitrary $N$-fermion density matrix. This would emphasize once
more that gravity is just an effective IR description of the
microphysics, and that it is pointless to take the metric
seriously once the Planck scale is reached. We leave such
explorations to future work.

%%%%%%%%%%%%%%%%%%%% START OF DISTRIBUTION FUNCTIONS %%%%%%%%%%%%%%%%

\subsection{Phase space distributions}

Above, we have discussed the analogy between (\ref{uDelta}), (\ref{uN}) and the phase space formulae for energies and populations of fermions in a harmonic potential.
In this section we shall explore in detail the exact quantum mechanical phase space description of half-BPS states in Yang-Mills theory.
In Section 4.3 we will  extract the semiclassical limit that is appropriate for comparing with the geometries of \cite{llm}, but the exact quantum structures described here will contain lessons about the origin of singularities in classical gravity.

\subsubsection{Wigner Distribution}

Given an $n$-dimensional quantum mechanical system described by the density matrix $\hat{\rho}_n$, the Wigner distribution function on phase space is defined by
\begin{equation}
  W(\vec{q};\vec{p}) =
  \frac{1}{(2\pi\hbar)^n}\,\int_{-\infty}^\infty\,d\vec{y}\
  \bra{\vec{q}-\vec{y}}\hat{\rho}_n\ket{\vec{q}+\vec{y}}\,
  e^{2i\,\vec{p}\cdot\vec{y}/\hbar}~.
\label{eq:defwigner}
\end{equation}
This function is real and its projections give probability densities on the configuration and momentum space:
\bea
&& \int d\vec{q}\ W(\vec{q},\vec{p}) = \bra{\vec{p}}\hat{\rho}_n\ket{\vec{p}}~, \\
&& \int d\vec{p}\ W(\vec{q},\vec{p}) = \bra{\vec{q}}\hat{\rho}_n\ket{\vec{q}}~, \\
&& \int d\vec{q}\ \int d\vec{p}\ W(\vec{q},\vec{p}) = {\rm tr}(\hat{\rho}_n)= n~.
\eea
Despite this, and except in special cases like Gaussian wavepackets, the Wigner function itself is not positive definite and is thus not in general a joint probability distribution on phase space.
Nevertheless, the Wigner function computes operator expectation values through the relation
\be
{\rm tr}(\hat{\rho}_n\,\hat{A}_W(\hat{\vec{q}},\hat{\vec{p}})) =
\int d\vec{q}\ d\vec{p}\ A(\vec{q},\vec{p})\ W(\vec{q},\vec{p})~,
\Label{Weylcorr}
\ee
where the operator $\hat{A}_W(\hat{\vec{q}},\hat{\vec{p}})$ is Weyl (symmetric) ordered in $\hat{\vec{q}}$ and $\hat{\vec{p}}$.
(Equivalently, for particles in a harmonic potential it is symmetric ordered in the ladder operators.)
The absolute value $|W(\vec{q},\vec{p})|$ is bounded.
In general the Wigner function oscillates rapidly inside the classical torus and decays exponentially outside of it. Useful reviews include \cite{wigner,berry,9204028}.

As we discussed in Section 2, the half-BPS states in $SU(N)$ Yang-Mills theory are related to the excited states of $N$ fermions in a harmonic potential.
The corresponding phase space is $2N$-dimensional.
However, the data relevant for the half-BPS solutions in gravity appear to involve only a two-dimensional phase space containing an effective single-particle distribution.
One computes such an effective one-particle distribution for an $N$-particle system by integrating out all but one of the particles from the $N$-particle density matrix:
\begin{equation}
  \rho_{1} = \int_{-\infty}^\infty dq_2 \, dq_3 \cdots dq_n\,\langle q_2,\ldots,q_n| \rho_n | q_2,\ldots,q_n \rangle =
  {\rm tr}_{{\cal H}_2 \otimes \ldots \otimes {\cal H}_n}(\rho_n)~.
\end{equation}
Though it may appear that we unnaturally singled out one of the
fermions in the above expression, due to the antisymmetry of the
$N$-fermion wavefunction $\rho_1$ is in fact completely symmetric
in all the fermions. We are interested here in the Wigner
distribution of a half-BPS state encoded by a single Young diagram
with row lengths $\CR = \{ r_1, \cdots r_N \}$. In the fermionic
representation of the state, the excitation levels are $\CF =
\{f_1 = r_1 , \cdots f_N = r_N + N-1\}$.\footnote{We have dropped
the irrelevant $\hbar/2$ zero-point energies.} Each fermion is in
an orthonormal pure state $(\Psi_{f_i}(q))$ and the wavefunction
of the $N$-fermion system is given by the Slater determinant of
the single particle wavefunctions
\be
\Psi(\vec{q}) = \frac{1}{\sqrt{N!}} \det\left( \Psi_{f_i}(q_j)
\right)~.
\ee
Due to the orthogonality of the $\Psi_{f}$, the single particle density matrix becomes
\begin{equation}
 \rho_1 (x,y) = \int d\vec{q}\ \Psi^*(x,\vec{q}) \Psi(y,\vec{q})
 = |C|^2\,\left(\sum_{f\in\CF}\,\psi^*_f(x)\,\psi_f(y)\right)\,,
\end{equation}
where $|C|^2$ is a normalization constant.
The effective single particle Wigner distribution function is defined as
\begin{equation}
W(q,p) = {1 \over 2\pi\hbar} \int_{-\infty}^\infty dy \, \langle q - y| \hat{\rho}_1|q + y \rangle e^{2i p y /\hbar}
\Label{effectivesingle}
\end{equation}

In our case, we are dealing with $N$ fermions in a harmonic potential.
The single particle wavefunctions are
\begin{equation}
\Psi_f(q) = A(f) H_f(q/\sqrt\hbar)\,e^{-q^2/2\hbar}~.
\end{equation}
$A(f)$ is a normalization factor and $H_f$ is a Hermite polynomial.
Following our general discussion, and using the identity
\begin{equation}
\int_{-\infty}^\infty dx\ e^{-x^2}\,H_m(x+y)\,H_n(x+z) = 2^n\,\sqrt{\pi}\,m!\,z^{n-m}\,L_m^{n-m}(-2yz), \quad m\leq n~,
\label{eq:iden1}
\end{equation}
the single particle Wigner distribution turns out to be
\begin{equation}
% u_W(q,p) = \frac{1}{\pi\hbar}\,e^{-(q^2+p^2)/\hbar} \sum_{n\in\CM}\,(-1)^n\,L_n\left(\frac{2}{\hbar}(q^2+p^2)\right)~,
W(q,p) = {1 \over \pi \hbar}\,e^{-(q^2+p^2)/\hbar} \sum_{f\in\CF}\,(-1)^f\,L_f\left(\frac{2}{\hbar}(q^2+p^2)\right)~,
\Label{eq:wignersea}
\end{equation}
where $L_n(x)$ is a Laguerre polynomial (the normalization is
$L_n(0) = 1$). The distribution is invariant under rotations in
the two-dimensional phase space as it should be for the states
under consideration.

%as $u_W(q,p)=W_1(q,p)$, we obtain
%\bea
%u_W(q,p) &=& \frac{|C|^2}{2\pi\hbar}\,2^{N(N-1)/2}\, \pi^{(N-1)/2}\,
%\prod_{n\in\CM} (n!)\,\int dx\,dy\,e^{-(x^2+y^2)/2}\, e^{-ip(x-y)} \nn \\
%         &&  \delta\left(\frac{x+y}{2}-q\right)\, \left(\sum_{n\in\CM}\,\frac{H_n(x)\,H_n(y)}{2^n\,n!}\right)~.
%\label{eq:genericwigner}
%\eea
%This can be evaluated exactly using the fact that $H_n(-x)=(-1)^n\,H_n(x)$, the new variables
%\begin{equation}
%w = \frac{x+y}{2}, \qquad
%z = \frac{x-y}{2} + ip~,
%\nonumber
%\end{equation}
%and the identity
%\begin{equation}
%\int_{-\infty}^\infty dx\ e^{-x^2}\,H_m(x+y)\,H_n(x+z) = 2^n\,\sqrt{\pi}\,m!\,z^{n-m}\,L_m^{n-m}(-2yz), \quad m\leq n~.
%\label{eq:iden1}
%\end{equation}
%The final result, after restoring $\hbar$ dependence that we have dropped in intermediate expressions, is
%\begin{equation}
%u_W(q,p) = 2\,e^{-(q^2+p^2)/\hbar} \sum_{n\in\CM}\,(-1)^n\,L_n\left(\frac{2}{\hbar}(q^2+p^2)\right)~,
%\label{eq:wignersea}
%\end{equation}
%where $L_n(x)$ is the Laguerre polynomial (the normalization is $L_n(0) = 1$).
%Notice that the resulting distribution is invariant under rotations in the two-dimensional phase space, in agreement with the subset of states under consideration.

\begin{figure}
  \begin{center}
     \epsfysize=2.0in
    \mbox{\epsfbox{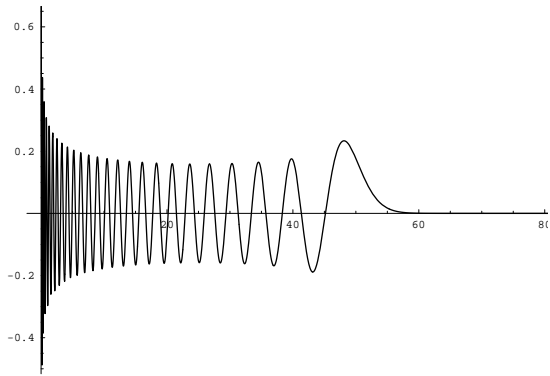}}
  \end{center}
 \caption{The Wigner distribution function of a single fermion with excitation level $n=50$ as a function of $H/\hbar$. }
\label{fig:wigner}
\end{figure}

\paragraph{A single excitation: }
Since the final distribution function is the sum of $N$ single particle Wigner distributions, we begin by considering the Wigner distribution of single fermion at excitation level $n$ \cite{groenewold}:
\be
W_{n}(q,p) = \frac{1}{\pi\hbar} (-1)^n e^{-2H/\hbar} L_n(4H/\hbar),
\qquad H(q,p) = \frac12 (p^2 + q^2)~. \\
\Label{singlefermionwigner}
\ee
At the origin of phase space, $W_n(0) = (-1)^n/\pi\hbar$.
As we move in the radial direction, the Wigner function oscillates with increasingly broad peaks (see Fig.~\ref{fig:wigner}).
The positions of the maxima and minima are given by the solutions to the equation
\be
\left(-\frac{\zeta}{2} + n \right) L_n(\zeta) - n L_{n-1}(\zeta) = 0,
\ee
and thus the oscillations occur at a scale $\sqrt{\hbar}$.
The final extremum is always a maximum and is positioned at $\zeta\approx 4n$, which coincides exactly with the classical orbit.
Beyond this maximum, the function decays exponentially.
Thus, the Wigner distribution is sensitive to fine quantum mechanical structures at the $\hbar$ scale.
In particular, the Wigner distribution for a single fermion is not well localized at scales below the classical energy.\footnote{
Well-localized Wigner distributions exist for coherent states which have a clean semiclassical limit.
These distributions describe the circular motion in phase space that we would intuitively expect for a classical oscillator.
All of this is in analogy with the fact that plane waves are completely delocalized in momentum whereas Gaussian wavepackets behave like classical particles.}  Indeed, it is not even positive.

Although the Wigner distribution is the most natural distribution in phase space \cite{berry}, the fineness of the quantum mechanical structure and the lack of positivity mean that it is not a good candidate for mapping into the boundary data required to determine the gravitational dual.
In particular to obtain the classical geometries described in \cite{llm}, we must ask how to take the appropriate semiclassical limit of the Wigner distribution.
The semiclassical limit requires $\hbar \to 0$.
To keep the Fermi level fixed we should take $N \to \infty$ in such a way that $\hbar N$ remains fixed.
Since energy levels are spaced by $\hbar$, phase space structures that do not vanish in this limit will involve $O(N)$ fermions at closely spaced energies.
From the Young diagram perspective, $O(N)$ rows with similar row length are necessary to give semiclassical structures.

%In the following we shall consider explicit examples of particular Young diagrams (the vacuum, a single excitation, the rectangular, and triangular diagrams), their corresponding exact Wigner distribution functions, and their semiclassical limits.
%Each of the Young diagrams is specified by a set of integers $r_n$, the number of boxes in the $n$-th row of the diagram, reading from bottom to top.
%The harmonic oscillators are excited to levels $r_n+n$.
%The extra $n$ is the energy of the fermion in the ground state.
%The corresponding Wigner distribution function for this state is
%given by
%\be
%u_W(q,p) = \sum_{n=0}^{N-1}\, W_{r_n + n}(q,p)~.
%\ee

\paragraph{The Fermi Sea: }
Consider placing the $N$ fermions in the $N$ lowest energy levels.
This corresponds to the vacuum of the theory.
There is no Young diagram describing the excitations because none of the fermions are excited ($r_n=0\,\,\forall\,n$ in the notation introduced above).
In the gravity description this should be empty $\AdS{5}\times S^5$, which is obtain by the boundary condition $u(0,r^2) = 1$ for $r^2/2 \leq E_f$ and $u(0,r^2) = 0$ for $r^2/2 > E_f$.
In the $N \to \infty$ limit the Wigner distribution is
\begin{equation}
2\pi\hbar W_{{\rm sea}}^\infty = 2\pi \hbar \sum_{n=0}^\infty
W_n(q,p) = 2 e^{-2H/\hbar} \sum_{n=0}^\infty (-1)^n L_n(4H/\hbar)
= 1, \Label{fermisea1}
\end{equation}
where $H = (p^2 + q^2)/2$, and we have used the summation formula
\be
\sum_{n\geq 0} L_n(x) z^n = (1-z)^{-1} \exp \left( \frac{xz}{z-1} \right) \, .
\Label{sumform}
\ee
We get the expected constant phase density for the vacuum, although the Fermi level has been pushed to infinity.
Taking $u(0, r^2) = W_{{\rm sea}} = 1$ to be the boundary condition in the $(x_1, x_2)$ plane does not lead to $\ads{5} \times S^5$ as a solution because the asymptotics of the geometry are changed.\footnote{
We thank Simon Ross for a discussion concerning this point.}
For any finite $N$
\be
2\pi\hbar W_{{\rm sea}} = 2\pi \hbar \sum_{n=0}^{N-1} W_n(q,p)
\ee
we find a function that has small oscillations around 1 for $(q^2
+ p^2)/2 < \hbar N$ and that approaches zero exponentially for
larger $E$. A more detailed discussion of this can be found in
\cite{berry}; here we will content ourselves by illustrating the point numerically. 
The scale for the decay and the oscillations is set by $\hbar$. In the strict limit
$\hbar \to 0$ with $\hbar N$ fixed, $2\pi \hbar W_{{\rm sea}}$
will be exactly one within the Fermi disk. Thus, in this limit we
should map $u(0,r^2) = 2\pi\hbar W_{{\rm sea}}(E)$ where $r^2 =
2E$. This is the familiar example of how the presence of many
fermions with adjacent energies produces a smooth semiclassical
distribution. We are also seeing that such a configuration in
field theory leads to a gravitational description with a smooth
semiclassical geometry. However, for any finite $N$ and $\hbar$ the Wigner distribution contains data on
$\hbar$ scale which maps onto the $\ell_p$ scale in gravity and is
not reflected in the classical vacuum geometry.

\begin{figure}
  \begin{center}
    \epsfysize=2.0in
    \mbox{\epsfbox{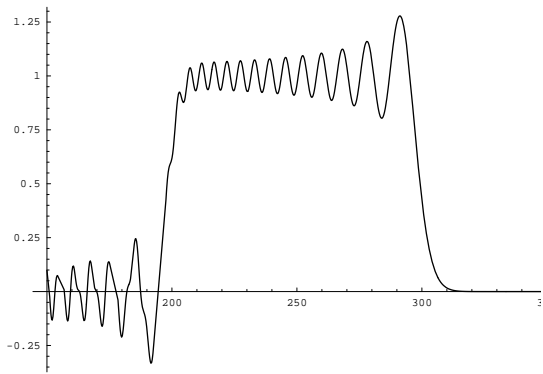}}
  \end{center}
 \caption{The sum of Wigner distribution functions of
single fermions with excitation levels between $n=200$ and $n=300$ as a function of $H/\hbar$. }
\label{fig:wignersumI}
\end{figure}

\paragraph{The rectangular diagram: }
Another state in which many fermions have closely spaced energies is described by the rectangular diagram, \eg
\begin{equation}
\Young[-8]{{40}{40}{40}{40}{40}{40}{40}{40}{40}{40}{40}{40}{40}{40}{40}{40}{40}{40}}
\Label{recttab}
\end{equation}
In such a state, each fermion has been excited by precisely the
same amount $\delta$ so that the operator creating the state has
conformal dimension $\Delta = N\delta$. In order for the energy
$\delta$ to be semiclassically visible we must take $\delta\hbar$
finite as $\hbar \to 0$. Since this is the same scaling as $N$ in
the semiclassical limit, we see that $\Delta \simeq N^2$. This is
satisfying because large classical objects in $\ads{5}$ always
have such large energies. Another way of saying this is that in
the semiclassical limit we rescale both dimensions of a Young
diagram by $1/N$ as $N \to \infty$ and only those diagrams that
survive this limit have a semiclassical description. The Wigner
distribution corresponding to (\ref{recttab}) is
\begin{equation}
2\pi\hbar W_{{\rm rect}} = 2\pi \hbar \sum_{n=\delta}^{N\delta} W_n(q,p) =
2\pi \hbar \sum_{n=0}^{N\delta} W_n(q,p) - 2\pi \hbar \sum_{n=0}^{\delta-1} W_n(q,p).
\end{equation}
Following the discussion of the Fermi sea, in the strict $\hbar
\to 0$, $N\hbar$ fixed limit, this will give a ring (for
$\delta\hbar \leq (p^2 + q^2)/2 \leq (N+1)\delta\hbar$) inside
which the Wigner distribution is equal to 1. Thus setting
$u(0;x_1,x_2) = W_{{\rm rect}}(q,p)$ in this limit reproduces the
``black ring'' boundary conditions described in \cite{llm}.
However, for any finite $N$ there are oscillations at the $\hbar$
scale both inside the inner radius of the ring and within the ring
itself (see Fig.~\ref{fig:wignersumI}). Again it is clear, that
the Wigner distribution cannot be directly mapped into gravity and
a prescription is needed for removing the quantum oscillations.

\paragraph{The triangular diagram: }
Consider a state characterized by a fixed gap of energy $\delta$ between any two fermions:
\begin{equation}
\Young[-6]{{40}{38}{36}{34}{32}{30}{28}{26}{24}{22}{20}{18}{16}{14}{12}{10}{8}{6}{4}{2}}
\Label{tritab}
\end{equation}
As discussed in Section 3.4, this is a typical state in an
ensemble where the total number of D-branes is $N_C$ and the
conformal dimension is $\Delta = N N_C/2$. Triangular Young
diagrams also provide the microscopic description of the superstar
geometry \cite{myers}, as will show in detail in
Section~\ref{starsss}. In the $N \to \infty$ limit, the Wigner
distribution of the triangular diagram is
\begin{equation}
2\pi\hbar W_{{\rm triangle}}^\infty = 2\pi \hbar \sum_{{k \geq 0}} W_{k\delta}(q,p)
= 2 e^{-2H/\hbar} \sum_{k=0}^\infty (-1)^{k\delta} L_{k\delta}(4H/\hbar)
\Label{triangleW1}
\end{equation}
with $H=(q^2 + p^2)/2$, the energy in phase space.
Next, summing (\ref{sumform}) over the roots of unity $\{\omega^0, \omega^1, \cdots, \omega^{(\delta - 1)/\delta} \}$ with $\omega = \exp(2\pi i/\delta)$ gives
\begin{equation}
\sum_{k\geq 0} (-1)^{k \delta } L_{k \delta }(4H/\hbar) = \frac{1}{\delta} \sum_{j=0}^{\delta-1} (1+\omega^j)^{-1} \exp \left( {4H \over \hbar} \frac{ \omega^j}{1+\omega^j} \right)
\Label{trianglesum}
\end{equation}
Notice that the real part of $\omega^j/(1+\omega^j)$ is always $1/2$.
Therefore all terms have the same overall $\exp(2H/\hbar)$ behavior.
The terms with $j\neq 0$ oscillate, over a distance of $\Delta H \sim O(\hbar)$.
Thus
\begin{equation}
 2\pi\hbar W^\infty_{{\rm triangle}} = {1 \over \delta} +  {\rm oscillations \ at \ scale \ \Delta H \sim \hbar }
\end{equation}
Even in the $N \to \infty$ limit, if $\hbar$ is finite, the oscillations persist, emphasizing that quantum oscillations in the Wigner distribution function cannot be erased simply by including many fermions.
This is different from the familiar case of the Fermi sea (\ref{fermisea1}) where the infinite sum erases quantum oscillations.
Notice however that if we coarse-grain this density over a distance larger than the typical wavelength of the oscillations, all terms with $j\neq 0$ will average to zero, and what remains is only the term with $j=0$.
We find a coarse-grained average density of $1/\delta$.

\begin{figure}
  \begin{center}
    \epsfysize=2.0in
    \mbox{\epsfbox{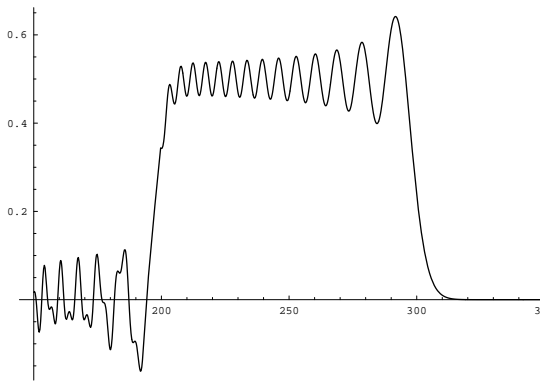}}
  \end{center}
 \caption{The sum of Wigner distribution functions of
single fermions with excitation levels between $n=200$ and $n=300$
separated by steps $\delta=2$ as a function of $H/\hbar$. }
\label{fig:wignersumII}
\end{figure}

Operators of any finite conformal dimension will of course correspond to diagrams for which the sum (\ref{triangleW1}) is truncated:
\begin{equation}
2\pi\hbar W_{{\rm triangle}} = 2\pi \hbar \sum_{k=0}^{M/\delta} W_{M_0+\delta k}(p,q).
\end{equation}
Numerically, one verifies that this Wigner function approaches a
homogeneous ring, with superimposed oscillations, that stretches
from $H=M_0$ to $M_0+M$ with density $u=1/\delta$, so long as $M$
is sufficiently large. Fig.~\ref{fig:wignersumII} plots the same
region as in Fig.~\ref{fig:wignersumI} but with $\delta=2$. We see
that the distribution in the classical region has its density
reduced by precisely this factor $\delta$.

The examples above show how the large-$N$ limit serves to build up
macroscopic distributions of fermions that can have a
semiclassical interpretation. Nevertheless, at any $N$ the phase
space distribution has important quantum mechanical features such
as oscillations, and excursions below zero and above one. Since
$\hbar$ in the field theory maps into $\ell_p$ in the bulk, we
will argue in Section 4.3 that this structure reflects an
underlying quantum mechanical structure in the spacetime at the
$\ell_p$ scale. But half-BPS spacetime solutions of \cite{llm} are
given in a supergravity theory that is effective in a
semiclassical limit that coarse-grains over distances much larger
than the Planck scale.   Therefore, as a first step towards this semiclassical limit, we will 
study the structure of the Wigner distribution coarse-grained at
the $\hbar$ scale and already find a structure that looks
considerably more classical than the Wigner distribution, though
it is still a complete quantum description of the system.

\subsubsection{Husimi distribution}
\begin{figure}
  \begin{center}
    \epsfysize=2.0in
    \mbox{\epsfbox{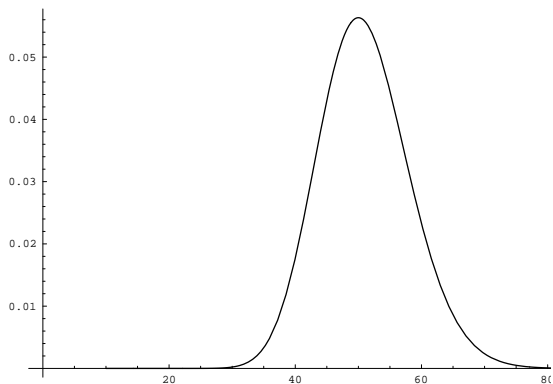}}
  \end{center}
 \caption{The Husimi distribution function of a
single fermion with excitation level $n=50$ as a function of
$H/\hbar$. Compare with figure~\ref{fig:wigner}. }
\label{fig:husimi}
\end{figure}

Given the $\hbar$ scale oscillatory behavior of the Wigner distribution function, it is natural to look for a second distribution function less sensitive to quantum mechanical structures, and more appropriate to the study of semiclassical physics.
One way of achieving this is by convolving the Wigner distribution itself with a Gaussian kernel of variance $\hbar$ to smooth out quantum oscillations.
For the effective single particle Wigner function $W(p,q)$ (\ref{effectivesingle}) we have
\bea
\Hu(q,p) &=& \frac{1}{2\pi\hbar} \int dQ\ dP\ Z(q,p;Q,P) \,W(Q,P), \label{trans} \\
Z(q,p;Q,P) &=& \exp\left[ - \frac{1}{2\hbar} \left( (q-Q)^2 + (p-P)^2 \right) \right].
\eea
The distribution function obtained in this way is called the
Husimi distribution $\Hu_n(q,p)$ \cite{husimi} and has several
other useful meanings. For example, for particles in a harmonic
potential, the Husimi distribution of a single particle state
$\ket\psi$ is a projection onto the coherent states $|z \rangle$:
\begin{eqnarray}
\Hu_\psi(q,p) &=& \frac{|\bra{z}\psi\rangle|^2}{\bra{z}z\rangle}~, \\
\ket{z} &=& e^{\bar{z} \hat{a}^\dagger - z \hat{a}} \ket{0} = e^{-|z|^2/2} \sum_n \frac{z^n}{\sqrt{n!}} \ket{n},
\end{eqnarray}
where $z=(p+iq)/\sqrt{2\hbar}$, $|n\rangle$ is the $n$th excited
state of the oscillator, and $\{ a,a^\dagger \}$ are the standard
ladder operators. For the eigenstates of the harmonic oscillator,
we have the Husimi distribution
\be
\Hu_n(z) = |\bra{z}n\rangle|^2 = \frac{1}{2\pi\hbar} \frac{1}{n!}
e^{-|z|^2} |z|^{2n}. \Label{Hun}
\ee
This is positive definite in
contrast to the corresponding Wigner distribution
((\ref{singlefermionwigner}) and Fig.~\ref{fig:wigner}). It has a
single maximum achieved at $|z|^2 = n$, which coincides with the
classical orbit of the $n$-th eigenstate since
$|z|^2=(p^2+q^2)/(2\hbar)$. (See Fig.~\ref{fig:husimi}.)
%Figure~\ref{fig:husimi} plots the Husimi distribution at level $n=50$.
The distribution function (\ref{Hun}) associated to single
particle eigenstates is also obtained as the wavefunction in
holomorphic quantization of the oscillator and as the wavefunction
in the associated quantum hall problem \cite{beren,shahin,davidrecent}.

Finally, just as the Wigner distribution function computes expectation values of Weyl ordered operators, the Husimi distribution computes those of reverse normal ordered operators $\hat{A}_N(\hat{q},\hat{p})$ in a state $\ket{\psi}$ through
\be
\bra{\psi}\hat{A}_N(\hat{q},\hat{p})\ket{\psi} = \int dq\ dp\ \Hu_\psi(q,p) A(q,p)~.
\Label{reversecorr}
\ee
This can be shown by using the identities
\begin{eqnarray}
(p -  iq)^n \langle \psi | e^{(p - iq) a^{\dagger}} | 0 \rangle
&=&
 \langle \psi | a^n  e^{(p - iq) a^{\dagger}} | 0 \rangle~, \\
 (p +  iq)^n \langle 0 |  e^{(p + iq) a} | \psi \rangle &=&
 \langle 0 |  e^{(p + iq) a} (a^{\dagger})^n| \psi \rangle~,
\end{eqnarray}
from which it follows that
\be
\int dp \,  dq \, \Hu_\psi(q,p) \,  (p-iq)^m \,  (p+iq)^n =
\langle \psi |  a^m (a^{\dagger})^n | \psi \rangle.
\ee
establishing the relation between the Husimi distribution and (reverse) normal ordered expectation values.

To study how peaked the maximum in the Husimi distribution is, we shall approximate the full distribution by a Gaussian, and compute its width as a function of the excitation level $n$.
Working with the variable $\zeta = (p^2+q^2)/(2\hbar)$, it is easy to show that for a single fermion the saddle point approximation gives
\begin{equation}
  \Hu_n(\zeta)= \frac{1}{2\pi\hbar}\,\frac{1}{n!}\,e^{-\zeta}\,\zeta^n
  \approx \frac{1}{2\pi\hbar}\,\frac{1}{\sqrt{2\pi n}}\,e^{-(\zeta-n)^2/2n}~.
  \Label{gaussapprox}
\end{equation}
The peak value is $\Hu_n(\zeta = |z|^2 =n)={1 \over (2\pi\hbar)}
{1 \over \sqrt{2\pi n}}$. Also ${\rm Var}(\zeta-n) \sim n$, and so
increases linearly with energy; the standard deviation grows as
$\sqrt{n}$. In terms of the physical energy variable $E = (p^2 +
q^2)/2 = \hbar \zeta$, the standard deviation is ${\rm \sigma}(E)
\sim \sqrt{n} \hbar \sim \sqrt{E} \hbar$. In terms of the radial
variable in phase space, $r = \sqrt{p^2 + q^2} = \sqrt{2E}$ this
translates into a standard deviation $\sigma(r) \sim
\sqrt{\hbar}$. In this sense, the Husimi distribution gives a
well-localized description of the harmonic oscillator eigenstates
in terms of rings in phase space whose width shrinks as $\hbar \to
0$. However, note that the spread in the variable $E$, $\sigma(E)
\sim \sqrt{E} \hbar$ is much greater than the separation between
adjacent energy levels $\Delta E = \hbar$. In this sense, the
Husimi distribution in phase space is still very far from being a
ring localized in a thin band at the classical energy --- the
classical ring approximation is simply wrong for these states.
While smearing the Wigner distribution with a Gaussian has removed
the small scale oscillations, the $\hbar$ scale quantum
uncertainties in the location and momentum of a single particle
are still very much present.

\begin{figure}
  \begin{center}
    \epsfysize=2.0in
    \mbox{\epsfbox{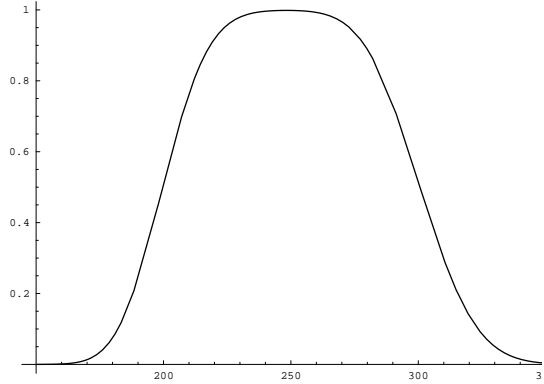}}
  \end{center}
 \caption{The sum of Husimi distribution functions of
single fermions with excitation levels between $n=200$ and $n=300$ as a function of $H/\hbar$.
Compare with figure~\ref{fig:wignersumI}. }
\label{fig:husimisumI}
\end{figure}

\begin{figure}
  \begin{center}
    \epsfysize=2.0in
    \mbox{\epsfbox{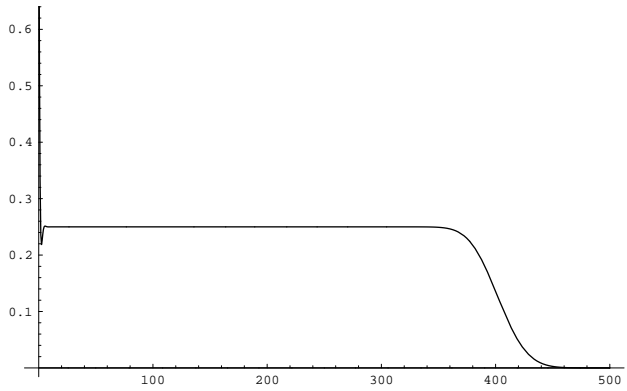}}
  \end{center}
 \caption{The sum of Husimi distribution functions for a triangular Young diagram; $\omega = N_C/N = 300/100 = 3$. The distribution plateaus at a grayscale of $1/(1+\omega) = 1/4$. }
\label{fig:husimi-triangle}
\end{figure}

We can also ask how the effective single particle Husimi
distribution behaves when we assemble many fermions together as we
must to obtain a semiclassical limit to compare with the half-BPS
spacetime geometries. The analysis here is much easier than it was
for the Wigner distribution because the answer is simply a
superposition of the well-localized functions (\ref{Hun}). To
assess how these superpositions behave let us study the $N \to
\infty$ limit of the Husimi distribution for the filled Fermi sea
and the triangular partition (\ref{tritab}). Using the identity
\begin{equation}
\sum_{n=0}^\infty {1 \over (\delta n)!} (\alpha x)^{\delta n} = {1 \over \delta} \sum_{k=0}^{\delta -1} e^{\alpha x\omega^k}  ~~~~~;~~~~~ \omega = e^{2\pi i/\delta}
\end{equation}
for $\delta = 1,2, \ldots$, it is easy to show that
\begin{equation}
2\pi\hbar \, \Hu^\infty_{{\rm triangle}} = 2\pi\hbar
\sum_{n=0}^\infty \Hu_{n\delta}(\zeta) = {1 \over \delta}
\sum_{k=0}^{\delta - 1} e^{- \zeta(1 - \exp(2\pi i k / \delta)) }
~~~~~;~~~~~ \zeta = {p^2 + q^2 \over 2 \hbar}.
\end{equation}
Thus in the $N\to \infty$ limit, the filled Fermi sea has a distribution
\begin{equation}
2\pi \hbar \,  \Hu_{{\rm sea}}  = 2\pi\hbar \, \Hu^\infty_1 = 1 \, .
\end{equation}
A triangular partition with row lengths increasing by $\delta - 1$ with $\delta$ large leads to
\begin{equation}
2\pi \hbar \,  \Hu_{{\rm triangle}}^\infty \approx {1 \over
\delta} \left[ 1 + 2 e^{- 4\pi^2 \zeta^2/\delta^2} \cos(2\pi
\zeta/\delta) + \cdots \right],
\end{equation}
where the additional terms are further suppressed by exponentials
of $\zeta$. It is easy to show that at the origin the Husimi
distribution for a triangular partition always approaches $1$,
and decays exponentially to a stable value of $1/\delta$ as
$\zeta$ increases. The rate of decay is controlled by $\delta$,
but for any $\delta$ of $O(1)$, the exponentially suppressed terms
are irrelevant for the $\zeta$ of $O(N)$ that are semiclassically
relevant. At any finite $N$ and $\hbar$, cutting off the sums at a
finite limit produces exponential tails at the edges of the
distribution that are determined by the variance in
(\ref{gaussapprox}). However, the oscillations that were present
in the Wigner distribution even after summing over many fermions
have been smoothed out (see Figs.~\ref{fig:husimisumI},
\ref{fig:husimi-triangle}). In all of these distributions, the
relatively large variance displayed in (\ref{gaussapprox}) is
important for producing a smooth distribution from the sum over
individual levels.

%%%%%%%%% TYPE NEW DISCUSSION %%%%%%%%%%%%%%%%%%%%%

\subsubsection{Semiclassical Limit: the Grayscale distribution}
\label{sec:gray}

We will now systematically derive the semiclassical limit of the Wigner and Husimi distributions that should have a relation to classical half-BPS spacetimes.
As we have described, the semiclassical limit requires taking $\hbar
\to 0$ with $\hbar N$ held constant\footnote{For a recent discussion
on the semiclassical limit of Wigner distribution function in the
context of non-minimal strings see \cite{cesar}.}.
In such a limit the Fermi energy, namely the energy of the highest fermion in the vacuum, is kept fixed.
Thus coordinates in phase space are effectively rescaled as $q \mapsto q \sqrt{\hbar}$ and $p \mapsto p \sqrt{\hbar}$.
In the language of integer partitions, the states surviving the semiclassical large-$N$ limit are those for which rescaling the lengths of rows and columns of the Young diagrams by $\hbar$ leads to a finite limiting shape.
In addition, quantum mechanical oscillations at the $\hbar$ scale should be averaged over;
this has been partially achieved by the Gaussian convolution of the Wigner distribution that led to the Husimi distribution.
To systematically derive a semiclassical distribution function on phase space, we should consider how semiclassical observables, namely operators that probe scales much larger than $\hbar$, will respond in a given quantum state.
One might worry that the ordering prescription for the operator will have a significant effect since the Wigner and Husimi distributions, computing the Weyl and normal ordered expectation values respectively, look so different from each other.
However, in the $\hbar \to 0$ limit the difference between ordering prescriptions vanishes: the semiclassical limit is universal.
Hence we will simply analyze the semiclassical limit of the Husimi distribution, as it is somewhat easier to handle.

We seek a new distribution function that is sufficient to describe the effective response of coarse-grained semiclassical observables in states that have a limit as $\hbar \to 0$.
For the single Young diagram states of interest here this distribution on phase space, which we name the {\it grayscale} distribution, will be only a function of the energy $E = (p^2 + q^2)/2$.
To derive this, let $\Delta E$ be a coarse-graining scale such that $\Delta E / \hbar \to \infty$ in the $\hbar \to 0$ limit.
This simply means that $\Delta E$ must vanish slower than $\hbar$ in the semiclassical limit.
A coarse-grained observable making measurements at an energy $E$ responds to the integrated fermion distribution in phase space between $E$ and $E + \Delta E$.
This response is given by
\begin{equation}
R(E,\Delta E) = 2\pi \hbar \, \int_E^{E + \Delta E} dq \, dp \, \Hu(q,p),
\Label{Req}
\end{equation}
where we have multiplied by $2\pi\hbar$ to account for the overall normalization of the Husimi distribution.
As discussed in the previous section, for a fermion of any fixed energy in the $\hbar \to 0$ limit, the Husimi distribution becomes a thin ring.
Therefore $R(E,\Delta E)$ counts the number of fermions with energies between $E$ and $E + \Delta E$.
To obtain the effective density we have to divide by the area in phase space in this region:
\begin{equation}
{\rm Area} = \int_{E}^{E+\Delta E} dq \, dp = 2\pi \Delta E
\Label{Areaeq}
\end{equation}
where we used $E = (p^2 + q^2)/2$.
Dividing (\ref{Req}) by (\ref{Areaeq}) gives the effective distribution that controls the response of observables coarse-grained at a scale $\Delta E$:
\begin{equation}
g(E) = 2\pi\hbar \left[ {R(E,\Delta E) \over  2 \pi \Delta E }\right].
\Label{gdef1}
\end{equation}
%In the language of the quantum Hall effect, this is the filling
%fraction for fermions in the lowest Landau level in the region
%specified by $E$ and $\Delta E$.
We can relate \eref{gdef1} to the
Young diagrams describing specific half-BPS states by recalling
that the number of fermions with energies between $E$ and $\Delta
E$ is simply given by the number of rows of the Young diagram with
the appropriate number of boxes. In terms of the coordinate $x$
introduced in Section 3 to index the rows of Young diagrams in the
large-$N$ limit,
\begin{equation}
g(E) = \hbar {\Delta x \over \Delta E} = {\hbar \over \partial E/\partial x}.
\Label{gdef2}
\end{equation}
Now consider any half-BPS state whose associated diagram is described by a limit curve $y(x)$ as obtained in Section 3.
Such a limit curve must exist for the state to have a well-defined semiclassical limit.
Otherwise, as $\hbar \to 0$ the structures in phase space become vanishingly small.
Given such a curve, the energy of the $x$th fermion is
\begin{equation}
E = \hbar (x + y(x))
\Label{energydef}
\end{equation}
Putting this together with (\ref{gdef2}) gives
\begin{equation}
g(E) = {1 \over 1 + y^\prime}
\end{equation}
where the right hand side is understood implicitly as a function of $E$ computed by inverting (\ref{energydef}) to find $x$ as a function of $E$.
This {\it grayscale distribution} captures the effective response of all coarse-grained semiclassical observables in the given quantum state.
Remarkably, this is precisely the quantity (\ref{eq:cmatch}) that we proposed on general grounds to determine the classical supergravity half-BPS solution associated to a Young diagram.
Thus, explicitly, our proposal is
\begin{equation}
u(0;r^2) = g(r^2/2) = {1 \over 1 + y^\prime}
\Label{proposal2}
\end{equation}
for all half-BPS states that are described by single Young
diagrams in the semiclassical limit.   To reiterate, we should
first determine $x$ as a function of $r=\sqrt{2E}$ from
(\ref{energydef}) and then substitute that function into
(\ref{proposal2}) to get the final form of $u(0;r^2)$.

\subsection{Quantum foam and effective singularity}
\label{starsss}

As we have discussed, half-BPS spacetime geometries are completely
specified by the values taken by a certain function
$u(\eta;x_1,x_2)$ on the $y=0$ plane. Non-singular, topologically
complex geometries result when $u(0;x_1,x_2)$ assumes the values
$0$ and $1$. Furthermore, an area quantization condition exists
that requires that compact droplets in which $u=1$ have their area
quantized in integral multiples of $2\pi \ell_p^4$ which was
identified with $\hbar$ in the field theory. These facts, along
with some explicit examples, suggested an identification between
the boundary values of the $u$ function in the $y=0$ plane and the
distribution of fermions in single-particle phase plane of the
dual fermionic picture of half-BPS states. Note, however, that the
classical solutions of \cite{llm} should be regarded as
parametrizing a  configuration space that still requires
quantization.  In particular in situations where the geometry
displays Planck scale features which have high curvatures such a
quantization is essential.\footnote{ Similar comments apply to the
non-singular solutions of \cite{Mathur,omm,mathura,simonvishnu,bena} in the D1-D5 setting.} An example is a
single droplet of unit area $2\pi\ell_p^4$. Our analysis has used
the dual field theory to find the exact quantum mechanical
structures describing each half-BPS state. In this section we will
argue that the exact structures imply an underlying ``foam'' at
the Planck scale whose effective description in low-energy gravity
is typically via various singular geometries. The singularity
emerges because underlying quantum structures have been integrated
out.

\paragraph{Wigner and ``quantum foam'': }
First, suppose we map the Wigner distribution $W$ associated to a
half-BPS state directly to gravity as $u = 2\pi\hbar W$. In
general this function is bounded, but can be negative and
certainly is not piecewise $0$ and $1$. Thus, a na\"{\i}ve map of
the Wigner distribution into classical gravity produces a singular
geometry, and, following \cite{caldarelli,milanesi}, the negative regions of
$W$ lead to tiny local closed timelike curves. However, both the
period of oscillations in $W$ and the size of the regions where
the function is negative, are set in field theory by the $\hbar$
scale and so translate in gravity into the Planck scale $\ell_p$.
Thus the na\"{\i}ve mapping from field theory to gravity produces
singularities at a scale where classical gravity should not be
trusted. A better interpretation is that the distribution is
describing an underlying exact quantum structure that is relevant
for the measurements made by certain kind of observables in
quantum gravity that are related to Weyl-ordered operators in
field theory. This underlying structure itself may not even have a
good description as an actual geometry. A typical operator,
encoded in a particular partition, has a detailed underlying
Wigner distribution with complex oscillations. In the fermionic
description the oscillations have two origins: (a) a regular
oscillation associated to each fermion, and (b) randomized
relative phases between contributions from different fermions,
arising from the random spacings between fermion energies. This is
describing an underlying ``quantum foam'' of extremely complex
structures at the Planck scale that do not have literal geometric
descriptions.

\paragraph{Husimi and ``classical foam'': }
An alternative proposal would be to map the Husimi distribution $\Hu$ directly into gravity.
This does not have quantum oscillations and is not negative, and builds up into coherent non-oscillating lumps when many fermions are piled up next to each other.
However, in general it is not equal to $0$ or $1$, as we have demonstrated in the previous section.
Thus, the effective geometry is singular.
A typical state corresponding to a particular partition would be a complex superposition of the Husimi lumps in Fig.~\ref{fig:husimi}, spaced apart in a random way.
If we artificially approximate these Husimi rings as 1s and 0s in thin rings, this will give a non-singular effective geometry with complex topological structures (different $S^3$s expanding) spaced randomly in the $x_1$--$x_2$ plane.
We might call this a {\it classical} foam.
However, because the typical structures will be at the $\hbar$ scale in the field theory, they will appear at the Planck scale in the geometry.
Hence, the curvatures will be large, and we cannot really trust the classical picture.
A better picture is that the non-singular geometries are a configuration space that should be quantized.
The exact Husimi distribution then describes the underlying quantum mechanical object which is being approximated by classical geometry.

We are in effect studying the quantum extension of the classical geometric considerations of \cite{llm} and \cite{Mathur}.
One might ask whether the Wigner or Husimi distribution is the correct extension, particularly given that they look so different.
However, the distributions in fact give complete descriptions of the same quantum wavefunction.
One is simply the natural description of measurements made by Weyl
ordered observables, while the other describes reverse normal ordering.
Indeed, each distribution can be transformed into the other.
The Husimi distribution is the convolution of a Gaussian with the Wigner distribution
\begin{equation}
\Hu(q,p) = \int dQ\,dP \, G(Q-q, P-p) \, W(q,p)
\Label{convolution}
\end{equation}
and this can be inverted as
\begin{equation}
W(k_q,k_p) = {1 \over G(k_q,k_p)} \Hu(k_q,k_p)
\Label{invert}
\end{equation}
where we have Fourier transformed the phase space variables $q$ and $p$ to turn the convolution (\ref{convolution}) into a product.
Using the definition of correlation functions (\ref{Weylcorr}) and
(\ref{reversecorr}) with these relations, it is easy to show that
reverse normal ordered operators that are non-local in phase space can recover that same information from the Husimi distribution that is encoded about Weyl-ordered operators encoded in the Wigner distribution.
The lesson about gravity is that the underlying ``foam''  looks different to different observables.
Local Weyl-observables see a more quantum mechanical structure, while local reverse-normal-observables see a more classical structure.
This sort of phenomenon has been seen before in string theory  --- D-branes, strings and gravitons can propagate in rather different effective geometries within the same spacetime.

\paragraph{Effective singularities: }
Of course a description in classical gravity presupposes that the semiclassical limit has been taken and that we are looking at observables that probe lengths much larger than the Planck length.
In the previous section we computed the phase space distribution (\ref{proposal2}) appropriate for such semiclassical observables for states that exist in the $\hbar \to 0$ limit.
We argued that this effective {\it grayscale distribution} is universal because the difference between Weyl and normal ordering prescriptions vanishes in the $\hbar \to 0$ limit.
One potential subtlety is that for operators of high dimension, commutators of many creation and annihilation operators can
give rise to c-numbers of order $N$, in which case there might be a finite correction even in the classical limit.
We will ignore this interesting possibility for the present.
We found that
\begin{equation}
{1 \over 2} - z = u = {1 \over 1 + y'(x)}.
\end{equation}
It is clear that states that do not correspond to rectangular, or step, diagrams ($y' \neq 0,\infty$) give boundary conditions in gravity that will lead to singular geometries.
The full geometry (\ref{solmetric}) is constructed in terms of a function
\begin{equation}
  z(\eta;r^2) = \eta^2\,\int_{\CD} dt\,
  z(0;t+r^2-\eta^2)\,\frac{t+2r^2}{[t^2+(2r\eta)^2]^{3/2}}~,
 \label{eq:generalz}
\end{equation}
This equation is obtained from (\ref{eq:zsol}) for rotationally
invariant functions $z(0;r^2)$ by performing the angular integral
and where the integration variable $t$ is related to $r'$, the original
radial variable in phase space, by $t = r'^2+\eta^2-r^2$. This
gives the universal low-energy spacetime description of quantum
states described by Young diagrams that lie close to the limit
curve $y(x)$. In this way, a singularity in  spacetime arises
because underlying quantum structure at the Planck scale has
been integrated out.

What information is lost in this semiclassical limit?
There are three obvious ways in which the description in classical geometry loses quantum detail:
\begin{enumerate}
\item The effective description of all states that lie close to a particular limit curve is the same.
Thus the detailed difference between these states has simply been lost.
For example, since almost all half-BPS states with a fixed total charge $\Delta$ lie close to the limit curve (\ref{eq:canlshape}), almost all such states will have the same description in classical gravity.
The differences between these states are encoded in gravity in Planck scale structures.
\item Given a particular microstate, the grayscale distribution appropriate to semiclassical observables (\ref{proposal2}) erases quantum details of the exact distribution on phase space.
These details are encoded in gravity in Planck scale structures that do not appear in an effective long-wavelength description in classical gravity.
The relevant Planck structures do not have a description in terms of smooth geometries.
\item The half-BPS quantum states are encoded in terms of a $2N$-dimensional $N$-fermion phase space, but only the effective two-dimensional one-particle phase space makes an appearance in the geometric description.
Some quantum observables, such as massive string states, could be sensitive to details of the full
$N$-particle phase space which has been erased.
\end{enumerate}
We will return to (3) in Section 5. Below we will discuss the
explicit construction of the effective geometries describing the
generic half-BPS states of fixed charge and number of D-branes
(which includes the well-known superstar \cite{myers}), and the
generic half-BPS states of fixed charge and unrestricted number of
D-branes (a new space that we will call the {\it hyperstar}).

\paragraph{The Hyperstar: }
Using the limit shape (\ref{eq:canlshape}) to compute the slope $y'(x)$, we can derive the grayscale by substituting into
(\ref{eq:cmatch}):
\begin{equation}
  z(0;r^2) = \frac{1}{2} - t\, q^{y(x)} \,\, \Leftrightarrow \,\,
  u(0;r^2) = t\, q^{y(x)} = 1 - q^{N-x},
 \nonumber
\end{equation}
where $t = q^{-C(\beta,N)}=1-q^N$.
In order to determine the full geometry, we need to relate the continuous variable $x$ with the radial variable $r^2$.
This can be done by using the identity $y(x) + x = r^2/(2\hbar)$, which we rewrite as
\begin{equation}
  \rho = \frac{r^2}{2\hbar} = C(\beta,N) + \frac{\log \left(1-q^{N-x}\right)}{\log q} + x~ \\ \Leftrightarrow \\
% x = \frac{\log \left(q^N + q^{-C(N)}\, q^{r^2/2\hbar}\right)}{\log q}.
  x = \frac{\log \left(q^N + t\, q^\rho\right)}{\log q}.
\nonumber
\end{equation}
The grayscale distribution in phase space is then 
\begin{equation}
  u(0;r^2) = \frac{e^{\beta\mu}\,e^{-\beta\,r^2/2\hbar}}
  {1 + e^{\beta\mu}\,e^{-\beta\,r^2/2\hbar}}~,
 \label{eq:matchhyper}
\end{equation}
which is nothing but a standard Fermi-Dirac distribution
with chemical potential
\begin{equation}
e^{\beta\mu} = q^{-N}-1
\end{equation}
also considered in \cite{buchel}.
The full type IIB configuration is therefore characterized by the function:
\begin{equation}
   z(\eta;r^2) = \frac{\eta^2}{2}\,\int_{\eta^2-r^2}^\infty dt\,
  \frac{1-e^{\beta\mu}\,e^{-\beta\,(t+r^2-\eta^2)/2\hbar}}{1+
  e^{\beta\mu}\,e^{-\beta\,(t+r^2-\eta^2)/2\hbar}}\,
  \frac{t+2r^2}{[t^2+(2r\eta)^2]^{3/2}}~.
 \label{eq:zhyper}
\end{equation}

Just as the limit shape (\ref{eq:canlshape}) has different behavior in different regimes, the same fact applies to the distribution function (\ref{eq:matchhyper}).
Consider the bottom of the Young diagram, that is, $(x/N)\ll 1$.
Since the limit shape is linear in this region, we know the dominant contribution to the distribution function should be a constant.
We shall also compute the first correction to this.
In this regime, the distribution function written in phase space variables is
\be
  u(0;r^2) = 1 - e^{-\alpha} - \alpha\,\frac{e^{-\alpha}}{\hbar\,N}
  \,(1-e^{-\alpha})\,\frac{r^2}{2}~.
 \label{eq:dropregimeI}
\ee
There is an upper bound on the radial coordinate $r^2$ due to the condition $x\ll N$ under which these expressions were derived.
This bound satisfies
\begin{equation}
  \frac{r^2}{2\hbar\,N} \ll \frac{1}{1-e^{-\alpha}}~.
\end{equation}
Thus, as expected, this is a regime in which we are describing the
phase space distribution at radial distances smaller than the scale
associated with the Fermi surface of the vacuum
$(\sqrt{\hbar\,N})$. We can replace $r^2$ with $t$ and integrate
this to determine the LLM function $z(\eta;r^2)$ in this regime.

Near the top of the Young diagram, that is, at $x\simeq N$, the grayscale distribution is approximated by
\begin{equation}
  u(0;r^2) = \alpha\,\left(1-\frac{x}{N}\right)~,
  \quad x \simeq N~.
\end{equation}
In this regime, the limit shape is a logarithmic curve, from which we can extract the relation between the radial coordinate in phase space and the row coordinate $x$:
\be
\frac{x}{N} = 1 - \frac{1}{\alpha}\ W\left(\frac{e^{-\alpha r^2/2N\hbar}}{e^\alpha(1 - e^{-\alpha})}\right),
\ee
where the right hand side is written in terms of Lambert's $W$-function,
which is obtained as the inverse function of the equation $W\ e^W = f(W)$.
The droplet distribution is
\begin{equation}
  u(0;r^2) = W\left(\frac{e^{-\alpha r^2/2N\hbar}}{e^\alpha(1 - e^{-\alpha})}\right).
\end{equation}
This regime probes scales in phase space much more energetic than the Fermi surface.

We should note an important consistency check to our analysis.
We have asserted that the conformal dimension is
\begin{equation}
  \Delta = \int_{\BR^2}
  \frac{d^2x}{2\pi\hbar}\ \frac{r^2}{2\hbar}\ u(0;r^2)
  - \frac{1}{2}\left(\int_{\BR^2} \frac{d^2x}{2\pi\hbar}\ u(0;r^2)\right)^2~,
  \label{eq:blah}
\end{equation}
which reproduces the expected results for semiclassical geometries which are non-singular.
This also applies to singular spacetimes like the hyperstar.
As the top row of the limit curve is infinitely long, we must integrate the excitation level $\rho$ of the fermions out to $\infty$.
We have
\be
\Delta = \left[ \rho \frac{\log( 1 + t\, q^{\rho-N} )}{\log q} + \frac{ \Li_2(-t\, q^{\rho-N})}{(\log q)^2} \right]_0^\infty -
         \frac12 \left( \left[ \frac{\log( q^N + t q^\rho )}{\log q} \right]_0^\infty \right)^2.
\ee
The second term gives
\be
\frac12 \left( N - \frac{\log( q^N + t )}{\log q} \right)^2.
\ee
As this term should fix the number of fermions (we are subtracting out the ground state energy corresponding to empty $\AdS{5}\times S^5$), we see that $t = 1 - q^N$.
Comparing to \eref{continuum}, we have indeed correctly reproduced $C(\beta,N)$.
Evaluating the first term as well, we now have
\be
\Delta = -\frac{\Li_2(1 - q^{-N})}{(\log q)^2} - \frac12 N^2 = \frac{\Li_2(1 - q^N)}{(\log q)^2},
\ee
where we have used a dilogarithm identity 
$
\Li_2(z) + \Li_2(\frac{z}{z-1}) + \frac12 \log(1-z)^2 = 0.
$
This is the expected result.\footnote{Since the distribution $u(0,r^2)$ has noncompact support, there is some question as to whether it describes an asymptotically $\ads{5} \times S^5$ geometry.  Of course $u$ falls off exponentially at large $r^2$, but AdS geometries also expand rapidly at large distances.}

\paragraph{The ensemble containing the Superstar: }
For a generic temperature, the limit shape describing a typical state in an ensemble in which the number of columns (D-branes) was fixed $(N_C)$ is given by equation (\ref{fsuperlshape}).
We can compute the slope $y'(x)$ of this curve at a given point, and from it, using our proposal (\ref{eq:cmatch}), derive the phase space distribution
\begin{equation}
z(0;r^2) = \left\{ \begin{array}{ll}
           \frac12 - (1-\zeta)q^{y(x)-N_C} & {\rm (within\ droplet)}, \cr
           \frac12                         & {\rm (outside\ droplet)}.
           \end{array} \right.
\end{equation}
Using the identity $y(x) + x = r^2/2\hbar$ and the limit shape (\ref{fsuperlshape}), we then invert $x$ as a function of $r^2$:
\begin{equation}
  \zeta\,q^{N-x} =
  \frac{1}{1+ e^{\beta\mu} q^{r^2/2\hbar}}~,
\end{equation}
where $e^{\beta\mu}$ again plays the r\^{o}le of a chemical potential,
\be
e^{\beta\mu} = q^{-(N+N_C)} \zeta^{-1} (1-\zeta) = \frac{q^{-N}-1}{1-q^{N_C}}.
\ee
The corresponding full type IIB configuration is determined by the integral expression:
\begin{equation}
   z(\eta;r^2) = \frac{\eta^2}{2}\,\int_{\eta^2-r^2}^{2\hbar(N+N_C)+\eta^2-r^2} dt\ 
  \frac{1-e^{\beta\mu} e^{-\beta\,(t+r^2-\eta^2)/2\hbar}}
  {1+e^{\beta\mu} e^{-\beta\,(t+r^2-\eta^2)/2\hbar}}
  \,
  \frac{t+2r^2}{[t^2+(2r\eta)^2]^{3/2}}~.
 \label{eq:zsuper}
\end{equation}
This assumes the same functional form as the corresponding equation for the hyperstar (\ref{eq:zhyper});
the differences are in the definition of the chemical potential and in the upper bound of the integration.
The upper bound is necessary because the droplet is compact.
There is no infinite tail to the Young diagram as in the hyperstar case.

As discussed in Section~3.4, in the infinite temperature limit the typical state is described by a triangular Young diagram.  In the large-$N$ limit this is described by a straight line limit curve (\ref{sline}).
The grayscale is a constant.
Using (\ref{eq:cmatch}), the phase space distribution $z_S(0;r^2)$ will also be constant:
\begin{equation}
  z_S(0;r^2) = \left\{ \begin{array}{ll}
               \frac12\, \frac{N_C/N - 1}{N_C/N + 1} & {\rm if}\ r^2/2\hbar \le N+N_C, \cr
               \frac12                               & {\rm if}\ r^2/2\hbar > N+N_C. 
               \end{array} \right.
 \label{eq:sdroplet}
\end{equation}
Since, within the droplet region, this number is different from $\pm 1/2$, the spacetime is again singular.

In the following we will show that the geometry derived from
(\ref{eq:sdroplet}) is precisely that of the
superstar \cite{myers}. This will be both a check of our
formalism, since we have already physically motivated that our
ensemble analysis is the appropriate one for describing the
microstates of the superstar, and also a confirmation of the
emergence of a singular spacetime as an effective description of microstates that differ from each other by Planck scale structures.

Let us first remember that the energy (conformal dimension) and total number of D-branes of the superstar configuration \cite{myers} are given by
\begin{equation}
  \Delta = \omega\,\frac{N^2}{2}~, \quad N_C = \omega\,N~,
  % \quad {\rm where}
  %\,\,\omega=\frac{q_1}{L^2}  ~,
\end{equation}
where $\omega = q_1/L^2$ is the parameter describing the charge of the system.
Notice that the slope of the limit shape (\ref{sline}) is precisely equal to $\omega$, \ie $y(x)=\omega\,x$.
It is now clear that the area under the limit shape equals the conformal dimension $(\Delta)$ given above.
Noting that
\be
u_S(0;r^2) = \frac12 - z_S(0;r^2) = \frac{1}{N_C/N+1},
\ee
in the region of the phase space plane between $r^2/2\hbar = 0$ and $r^2/2\hbar = N + N_C$, we may also readily verify from \eref{eq:blah} that the distribution function also reproduces $\Delta$.
The quantized number of giant gravitons $N_C$ is encoded in the radius of the gray droplet, or as the integral of $(1-u_S)$ over the droplet. 

Let us rewrite the ten-dimensional metric corresponding to the half-BPS superstar configuration in type IIB:
\begin{eqnarray}
  ds^2 &=& -\frac{\sqrt{\gamma}}{H_1}\,f\,dt^2 +
  \frac{\sqrt{\gamma}}{f}\, dr^2 + \sqrt{\gamma}\,r^2\,ds^2_{S^3} +
  \sqrt{\gamma}\,L^2\,d\theta_1^2 + \frac{L^2}{\sqrt{\gamma}}\,
  \sin^2\theta_1\,ds^2_{\tilde{S}^3} \nonumber \\
  & & + \frac{H_1}{\sqrt{\gamma}}\cos^2\theta_1\,[
  L\,d\phi_1 + (H_1^{-1}-1)\,dt]^2~,
 \label{eq:ssmetric}
\end{eqnarray}
where $H_1=1+q_1/r^2$, $f=1+r^2\,H_1/L^2$ and $\gamma=1+q_1\sin^2\theta_1/r^2$.
If we compare the physical size of the two three-spheres that appear in the superstar metric with their parametrization in \cite{llm}, we obtain the conditions:
\begin{eqnarray}
  \eta\,e^G &= & \sqrt{\gamma}\,r^2~, \nonumber \\
  \eta\,e^{-G} &= & \frac{L^2\,\sin^2\theta_1}{\sqrt{\gamma}}~.
\end{eqnarray}
Solving the system
\begin{equation}
  \eta^2 = r^2\,L^2\,\sin^2\theta_1~, \quad
  e^{G} = \frac{r\sqrt{\gamma}}{L\sin\theta_1}~,
 \nonumber
\end{equation}
and using the fact that the grayscale distribution is given by
$z=(1/2)\tanh G$ we obtain:
\begin{equation}
  z = \frac{1}{2}\frac{r^2\gamma - L^2\sin^2\theta_1}
  {r^2\gamma + L^2\sin^2\theta_1}~.
\end{equation}
Since it is the value $z(\eta=0)$ that is related to the semiclassical distribution function, we must analyze the behavior of $G$ at $\eta=0$.
We observe that there are two different coordinate regimes for which this applies:
\begin{itemize}
  \item[(i)] When $\sin\theta_1=0$, $z(\eta=0)=1/2$.
  This is consistent with the fact that whenever $\sin\theta_1$ vanishes, the distribution of D-branes vanishes.
  There are no fermion excitations in this locus.
  Since we follow the conventions in \cite{llm}, $1/2$ is the right boundary value to describe absence of excitations in the fermionic picture.
  \item[(ii)] When $r=0$, the density of giants is non-vanishing.
  We get
\begin{equation}
  z(r=0) = \frac{1}{2}\frac{\omega -1}{\omega +1}~.
 \label{eq:zgiant}
\end{equation}
\end{itemize}
  It is reassuring to check that the expression (\ref{eq:zgiant}) identically matches what is derived from purely field theoretic and statistical mechanical considerations (\ref{eq:sdroplet}).

\section{Correlation functions}

In Section~2.2 we argued that correlation functions computed in typical very heavy states are to
great accuracy independent of both the operators that appear in
the correlation function as well as the state that is being
probed. Correlation functions therefore only depend on charges and
other quantum numbers that clearly distinguish states and
operators at the macroscopic level. Deviations from universality
are expected to be of the order of $\exp(-S)$, with $S$ the
entropy of the ensemble from which we pick the states that we
probe. In this section we will try to make these claims somewhat
more precise for the case of half-BPS states.

In general, correlation functions computed in a half-BPS state will
involve the full Yang-Mills theory because the intermediate states in
an arbitrary correlation function can explore the full theory.
However, there are special classes of correlation functions which can
be computed entirely within the half-BPS matrix models.  The authors
of \cite{CJR,koch} describe such correlators and group theory
techniques for computing them.     Here we will take a different
perspective and simply use the Wigner (or Husimi) distribution
function to compute correlation functions as phase space integrals.
 While we will will not compute the general correlation function of the
theory, it is nevertheless interesting to ask whether these matrix
model correlators are sensitive to the difference between typical states.  In general we expect our results to  
be renormalized in the full theory, but the matrix model results give useful insights anyway.  It would be interesting to work out the representation of the operators with exact matrix model correlators in the language of functions on phase space.

\subsection{Phase space approach}

The natural operators in the half-BPS matrix model are products of
${\rm tr}(X^p)$. Such operators can be rewritten in terms of
${\rm tr}(A^p)$ and ${\rm tr}((A^{\dagger})^q)$, the matrix
creation and annihilation operators. By definition, all
correlation functions computed within the
half-BPS matrix model can be equivalently calculated through phase space
integrals of the corresponding phase space distribution functions.
Generic multiple trace operators involve $2N$-dimensional phase space
integrals, and so they require the knowledge of the $2N$-dimensional
phase space distribution.  On the other hand, the semiclassical
gravitational description of any typical state in the ensemble only
depends on the $2$-dimensional single-particle phase space
distribution function, raising the question of what data about the
full quantum mechanical system survives the semiclassical limit.

The special class of single-trace operators has expectation values
that can be computed by just knowing the single-particle phase space
distribution. Such operators, when expressed in terms of eigenvalues,
become a sum $\sum_{i=1}^N F(\lambda_i,\frac{\partial}{\partial
\lambda_i})$ with $\lambda_i$ the i${}^{\rm th}$ eigenvalue of $X$.
This is a sum of one-fermion operators, and thus its expectation value
can be completely computed using the single-particle Husimi or Wigner
distribution $u(q,p)$. We will focus our attention on such operators here.

We are interested in computing the expectation values of single trace
operators in states appearing in certain ensembles, as well in
measuring the variance over the ensemble of these responses.
The variance to mean ratio will give a measure of the universality of
correlation functions computed in typical states.  Let us denote
by $u_{\lambda}(\vx)$ the Wigner or Husimi distribution of the
partition $\lambda$ (state), where $\vx = (q,p)$ stands for the phase
space point. The ensemble average of the distribution function
$u_w(\vx)$ in an ensemble ${\cal E}$ with weight $w(\lambda)$ will be given by
$u_w(\vx) = \sum_{\lambda \in {\cal E}} w(\lambda) u_\lambda(\vx)$.
By definition, the expectation values of observables $F$ and their
corresponding ensemble averages are
\begin{eqnarray}
F_\lambda &=& \int d\vx \, F(\vx) \, u_\lambda(\vx) \, \Label{obsave}\\
F_w &=& \int d\vx \, F(\vx) \, u_w(\vx) = \langle F_\lambda
\rangle_w~, \Label{ensave}
\end{eqnarray}
respectively. On the other hand, the variance in this observable over the
ensemble is
\begin{equation}
{\rm Var}_w(F) =
\sum_{\lambda \in {\cal E}} w(\lambda) \left[F_\lambda
- \langle F_\lambda \rangle_w \right]^2 \, .
\Label{varobsdef}
\end{equation}
Using the definitions for $F_\lambda$ and $\langle F_\lambda
\rangle_w$, this variance can be written as
\begin{equation}
{\rm Var}_w(F) =
\int d\vx \, d\vy \, F(\vx) \, F(\vy) \, \sum_{\lambda \in {\cal E}} w(\lambda)
\left[ u_\lambda(\vx) - u_w(\vx) \right] \left[ u_\lambda(\vy) - u_w(\vy) \right] ~.
\Label{varobs}
\end{equation}
As expected, this is an expression that depends on the fluctuations of
$u_\lambda(\vx)$ around its mean ensemble value $u_w(\vx)$ at
different points $\vx$.

For the subset of states considered in this paper, all phase space
distributions are rotationally invariant, and so we write them as
$u(r)$, $r$ being the radial variable in phase space. Furthermore, from our
canonical ensemble analysis, the fluctuations in the number of Young
diagram columns of different lengths are uncorrelated. It is easily
shown that this translates in phase space into the independence of
the fluctuations of $u_\lambda(r)$ around its mean ensemble value $u_w(r)$ at
different points $r$.   In other words,
\begin{equation}
\sum_{\lambda \in {\cal E}}w(\lambda) \left[ u_\lambda(r) - u_w(r) \right]
\left[ u_\lambda(r') - u_w(r') \right] =
\delta(r - r') \sum_{\lambda \in {\cal E}}
w(\lambda) \left[ u_\lambda(r) - u_w(r') \right]^2~.
\label{nocorr}
\end{equation}
Using this, we can rewrite (\ref{varobs}) as
\begin{equation} \label{sigsig}
{\rm Var}_w(F) =
\int d\vx \, F^2(r) \left[ \langle u_\lambda^2 \rangle_w
- \langle u_\lambda\rangle_w^2 \right]~,
\end{equation}
where we also assumed for simplicity that the operator $F$ itself was rotationally
invariant. Therefore we learn that the variance in observables over
the ensemble ${\cal E}$ is controlled by the variance of the phase
space distribution over the same ensemble.

Let us first consider very delocalized observables spread over the
entire compact region where the distribution $u_\lambda$ has support.
(Of course $u_\lambda$ may have exponential tails, but these will not
make a large contribution, so we can ignore them and speak about the
region of support as being compact.)   Then for any reasonable
observable $F$ can be bounded by some $F_0$ within the region of
support and we can bound the mean and variance of the observable by
\begin{eqnarray}
F_w &\leq& F_0 \int d\vx \, u_w(\vx) = F_0 N \,, \\
{\rm Var}_w(F) &\leq& F_0^2 \int d\vx \,\langle u_\lambda^2 \rangle_w
\leq F_0^2 \int d\vx \, u_w(\vx)  = F_0^2\, N\,.
\end{eqnarray}
For sufficiently delocalized observables this leads to the estimate
\begin{equation}
{{\rm Var}_w(F) \over F_w^2} \sim {1 \over N}
\end{equation}
so that deviations from universality behave
like $e^{-N}$. For most ensembles we consider the entropy scales
as $\sqrt{\Delta}\sim N$, so this is in agreement with our general
considerations in Section~2.2.

Next consider a highly localized observable $F(q,p) \sim F_0 \,
\delta(\sqrt{q^2 + p^2} - r_0)$. In fact, such a probe localized
below the $\hbar$ scale is unrealistic, but it is instructive to
study it anyway.  In this case
\begin{eqnarray}
F_w &\sim& F_0 \, \langle u_\lambda(r_0)\rangle_w\, , \\
{\rm Var}_w(F) &\sim& F_0^2 \, \left[\langle u_\lambda^2(r_0)
\rangle_w -  \langle u_\lambda(r_0)
\rangle_w^2 \right]
= F_0^2 \, {\rm Var}_w(u_\lambda(r_0))
\, .
\end{eqnarray}
In the semiclassical limit, both the Wigner and the Husimi
distributions approach the grayscale distribution of Section~4.2.3 and so
in this limit we take $u_w(r^2) = 1/(1 + y')$ in terms of $y(x)$, the
Young diagram limit shape.\footnote{Strictly speaking, in the semiclassical limit we should not be using observables that are localized at scales smaller than $\hbar$. But we will do this anyway to get a sense of the effects of localizing a probe in phase space.} Using standard error propagation, it easy
to relate the variance in $u$ for small $r^2$, which is the region
containing most of the structure of half-BPS states,\footnote{In the small $r^2$, the fluctuations are small, but there
are enormously many configurations.  At large $r^2$ the fluctuations
are larger and the error propagation formula is not really valid, but
there are also many fewer configurations to consider.}
to the variance in the $y^\prime(x)$. Doing this one shows using (\ref{varcont}) that
\begin{equation}
{{\rm Var}_w(F) \over F_w^2}
= {\rm Var}_w(y^\prime) \, u_w^2(r_0)
\Label{deltavar}
\end{equation}
Finally, by definition (\ref{continuum})
\begin{equation}
y^\prime = {\int_{N-(x + \Delta x)} ^N di \, c_i  - \int_{N-x} ^N di \, c_i   \over \Delta x} \approx c_{N-x}
\end{equation}
where $c_{N-x}$ is the number of columns of length $N-x$ in the Young diagram. From this,
\begin{equation}
{\rm Var}_w(y^\prime(x)) = {\rm Var}_w(c_{N-x}) = {e^{-\beta (N - x)} \over (1 - e^{-\beta (N - x)})^2}
\Label{yprimevar}
\end{equation}
Putting this together with (\ref{deltavar}) and recalling that $\beta$ is $O(1/N)$ for the ensembles of interest we see that for small $x$ (near the origin in phase space) where $u_w \sim 1$, the delta function observable has $O(1)$ variations over the ensemble.   For $x$ of $O(N)$ (where the error propagation formula we used is not strictly valid) the enhancement of the variance by the denominator in (\ref{yprimevar}) competes with the exponential decay in $u_w$.    In any case, the main point is that extremely localized observables show larger variation over the ensemble and thus can be used to identify details of the state.  However, the $\delta$ function observable considered here will be extraordinarily difficult to construct as a probe in the dual gravity theory. Recall that the $\hbar$ scale in the field theory maps into the Planck scale in gravity, so we are describing the results of measurements by a super-Planckian probe.   Such things will be extremely difficult to construct in gravity, because gravitons, strings, D-branes and black holes, which are the natural probes in string theory, all have sizes much bigger than the Planck length.  Indeed, in general adding energy to a probe in string theory tends to increase its size.  In fact, to construct probes that measure super-Planckian scales, non-local objects may be needed.   To see this, observe that  in the dual field theory that we are discussing here, the natural probes are polynomials in $X$ with traces distributed to impose gauge invariance.  These objects map into the usual gravitational probes.    Local probes in the field theory simply {\it cannot} be localized below the $\hbar$ scale according to the Heisenberg uncertainty principle.   Thus,   one is led to consider non-local combinations of objects involving arbitrarily large frequencies.

A particularly efficient set of observables $\{ F_n(q,p) \}$ would
be functions of phase space that are the single fermion
distribution functions.  This is because the full phase space
distribution is simply a linear sum of such constituent terms.
Thus integrating the $F_n(q,p)$ against the distribution function
for an individual diagram $\lambda$ acts as a filter for testing
whether the associated state contains the excitation level $n$. In
this way it would appear that there is an $O(N)$ set of
observables that can efficiently ``detect'' the half-BPS state
associated to a given Young diagram.   This phenomenon is
occurring because the half-BPS states are an integrable sector of
the full Yang-Mills theory.  This leads to the free nature of the
system and corresponding simplicity of identifying states.
However, note that the filtering observables $\{ F_n(q,p) \}$ will
not be simple to construct in terms of the  multi-trace
observables that are naturally related to probes in gravity.   We
should expect that they are related to non-local observables of
the Yang-Mills theory like other integrable charges of this system
\cite{integrable}.   Nevertheless, the basic idea here that
certain probes extract patterns from the underlying quantum state
is in analogy to the discussion of Schwarzschild black holes in
Section~2.

Another approach to computing correlation functions using the fermion phase space is given in Appendix A.

\paragraph{Exact correlation functions in phase space: } Since we
know the exact single particle quantum mechanical Wigner distribution
function for a given  Young diagram, it is possible to compute
the exact matrix model correlators for single particle
operators. To be definite, let us consider some observable whose phase
space description is given in terms of $p^t\,q^s$, i.e.
$F(p,q;t,s)=p^t\,q^s$.     We wish to compute the expectation of the Weyl ordered operator corresponding to $F(p,q;t,s)$ in a state corresponding to a partition $\lambda$.    From Section~4, the Wigner distribution of the state is $W_\lambda = \frac{1}{\pi\hbar}\,e^{-(p^2+q^2)}\,\sum_{m\in\lambda}\,(-1)^m\,
L_m[2(p^2+q^2)]$.  Working with polar coordinates in phase space, the
expectation value is then given by
\begin{equation}
  F_\lambda(t,s) = \int_0^\infty rdr\,\int_0^{2\pi}d\phi\,r^{t+s}\,
  \sin^t\phi\,\cos^s\phi\,W_\lambda~.
\end{equation}
Due to rotational invariance of the state, the above integral is only
non-vanishing when both $(t,s)$ are even $(t,s)=(2\ell,2n)$. Using this
fact, and some identities, the result is
as
\begin{equation}
  F_\lambda(t,s) =
  \delta_{t,2\ell}\,\delta_{s,2n}\,\hbar^{n+\ell-1}\,
  \frac{(2n-1)!!(2\ell-1)!!}{2^{n+\ell}}\,\sum_{m}\,(-1)^m\,F(-m,\ell+n+1,1;2)~.
  \Label{exactcorr}
\end{equation}
A similar exact expression can be given for normal ordered correlators using the Husimi distribution.
From this point of view, the question of identifying the partition $\lambda$ from the set of observables $F(p,q; t,s)$ amounts to  asking how much detail in (\ref{exactcorr}) is needed to tell apart correlations in two different typical states.  This might be a convenient formulation of the problem.    It would also be useful to translate the functions $F(p,q;t,s)$ back into the language of gauge invariant operators of the Yang-Mills theory in order to identify precisely what observables these are in string theory.

\paragraph{Multi-particle phase space: }If $F$ is not a sum of one-fermion operators, for example if we
are interested in computing the correlation function of ${\rm
tr}(M^k) {\rm tr}(M^l)$ in the matrix model, the result can no
longer be written  as a simple integral over the single particle phase space. For
ensembles like (\ref{obsave}), one can show that expectation values of
operators that can be written as sums of $t$-fermion operators
reduce in general to $t$-fold integrals over phase space, with
rather complicated kernels that are functions of the $t$-points.
This is quite natural from the point of view of the AdS/CFT
correspondence, where $t$-point functions involve multiple
integrals over the bulk as well. In the present case the bulk is
in general singular, but we nevertheless find concrete and precise
expressions for the multipoint correlators as multiple integrals
over phase space. Thus, in this half-BPS context, we have in principle a
precise proposal for the boundary conditions that need to be
imposed at the singularities. It would be interesting to
explore this further.

\subsection{Second quantized approach}

We can try to improve upon the above analysis in a second quantized formalism.
Consider  a free Fermi field with a Hilbert space with
modes $\psi_m$, $m\in \mathbb Z + \frac{1}{2}$, and with
commutation relations $\{ \psi_m,\psi_n \} = \delta_{m+n}$.  We will restrict attention to the $N$-fermion sector of this Hilbert space.   The
action of the matrix creation and annihilation operators is given
by the following fermion bilinears
\bea
{\rm tr} (A^p) & \leftrightarrow & \sum_{n\geq 0}
\sqrt{\frac{(n+p)!}{n!}} \psi_{-\frac{1}{2}-n-p}
\psi_{\frac{1}{2}+n} \nonumber \\
 {\rm tr} ((A^{\dagger})^p) &
\leftrightarrow & \sum_{n\geq 0} \sqrt{\frac{(n+p)!}{n!}}
\psi_{-\frac{1}{2}-n} \psi_{\frac{1}{2}+n+p} \label{fermops}
\eea
and one can in a similar way work out more complicated single
trace operators that consist of a product of terms $A^p$ and
$(A^{\dagger})^q$.

To illustrate the computation of correlation functions, consider
\be {\cal O}={\rm tr} (A^p) {\rm tr}
((A^{\dagger})^p).
\ee
For quantities like (\ref{obsave}), we only need the piece of ${\cal
O}$ that acts diagonally on the basis of states consisting of
partitions $\lambda$. If we insert (\ref{fermops}) in ${\cal O}$,
the only contributions to (\ref{obsave}) are therefore from
\be
{\cal O}_0 =\sum_{n \geq 0} \frac{(n+p)!}{n!}
\psi_{-\frac{1}{2}-n} \psi_{\frac{1}{2}+p+n}
\psi_{-\frac{1}{2}-p-n} \psi_{\frac{1}{2}+n}.
\Label{y4}
\ee
Such an operator gets contributions only from fermionic states
with both the $n$-th and $(n+p)-$th harmonic
oscillator levels are occupied. When computing the ensemble average
(\ref{ensave}), it is natural to expect to get the same contributions
as above, but weighted with the probability that such excitations
occur in the ensemble. Since in this paper we have only considered
thermal fermionic distributions, these probabilities should be given
by the standard Fermi-Dirac distribution:\footnote{This is consistent
with the fact that, in the semiclassical limit, the phase space
distribution obtained through our analysis of the limit shape Young
diagram is precisely given by the Fermi-Dirac distribution.}
\be
v(n) = \frac{q^{n-\mu}}{1+q^{n-\mu}}~.
 \Label{y6}
\ee
Thus the ensemble average (\ref{ensave}) of ${\cal O}$ will be equal to
\be \label{y7}
\langle {\cal O} \rangle_w = \sum_{n \geq 0} \frac{(n+p)!}{n!} v(n)
v(n+p).
\ee
For small values of $p$ this can be approximated as follows
\bea
\langle {\cal O} \rangle_w & \sim & \sum_{n \geq 0} n^p v(n)^2
\nonumber \\
& \sim & \int_0^{\infty} \, dn\, n^p v(n)^2.
 \Label{y8}
\eea
More generally, we can consider integrals $\int \, dn\, n^{\alpha}
v(n)^{\beta}$. If we expand the denominator in $v(n)$ and do the
$n$-integral, this reduces to
\bea
\int_0^{\infty} \, dn\, n^{\alpha} v(n)^{\beta} & =  &
\Gamma(\alpha+1) \left( \frac{-1}{\log q} \right)^{\alpha+1}
\sum_{r\geq 0} (\beta+r)^{-\alpha-1} \left( \begin{array}{c}
-\beta \\ r
\end{array} \right) q^{-\mu(\beta+r)} \nonumber \\ & \sim &
\Gamma(\alpha+1) \left( \frac{-1}{\log q} \right)^{\alpha+1}
\sum_{r\geq 0} \frac{(-1)^r}{(\beta-1)!} q^{-\mu(\beta+r)}
(1+r)^{\beta-\alpha-2} \nonumber \\
& = & -\Gamma(\alpha+1) \left( \frac{-1}{\log q}
\right)^{\alpha+1} \frac{q^{-\mu(\beta-1)}}{(\beta-1)!} {\rm
Li}_{2+\alpha-\beta}(-q^{-\mu}).
 \Label{y9}
\eea
In particular, turning back to (\ref{y8}) we get
\be \label{y10}
\langle {\cal O} \rangle_w \sim -\Gamma(p+1) \left( \frac{-1}{\log
q} \right)^{p+1} q^{-\mu} {\rm Li}_{p}(q^{-\mu}).
\ee
There are a couple of interesting features in this expression.
First, its scaling with $N$ is controlled by $(-1/\log q)^{-p-1}$.
Second, as we claimed in Section~2.2, the ensemble average does
only depend on the parameters specifying the ensemble
$\{\mu,\,\beta\}$. These are in one-to-one correspondence with the
global charges $\{N,\,\Delta\}$ characterizing the set of
microstates under study. Thus we see that once we average over the
ensemble, which is the right way to infer the properties of
correlation functions in typical states, the response is mainly
determined by their global charges.

Let us compute the variance in the expectation value of ${\cal O}$ to
estimate the universality of the above result as we consider different
typical states. The only contributions to (\ref{varobsdef}) from
${\cal O}$ are coming from the diagonal part of ${\cal O}$ which was
explicitly written down in (\ref{y4}) and so it equals
\be
{\rm Var}_\lambda ({\cal O}) = \langle ({\cal O}_0)^2 \rangle - \langle {\cal O}_0
\rangle^2.
\Label{y11}
\ee
Just as the operator ${\cal O}_0$ gets contributions if the harmonic
oscillator eigenstates $n$ and $n+p$ are occupied, $({\cal O}_0)^2$
gets contributions if eigenstates $n$, $n+p$, $m$ and $m+p$ are
occupied for some $m,n$. Generically these are four
different eigenstates, but when $|m-n|=p$ there are only three and
if $|m-n|=0$ only two. These special configurations of $m,n$ are
the only ones that contribute to (\ref{y11}) in thermal ensembles.
All other (generic) configurations of $n,m$ are uncorrelated and
cancel against each other in the two terms in (\ref{y11}).
As in (\ref{y7}), when we compute the ensemble variance, we weight
these contributions with the probabilities for such harmonic potential
excitations to occur in the ensemble. This way, the ensemble variance is
\bea \label{y12}
{\rm Var}_w ({\cal O})  & = & 2\sum_{n\geq 0} \frac{(n+2p)!}{n!} v(n) v(n+p)
v(n+2p) (1-v(n+p)) \nonumber \\
& & + \sum_{n \geq 0} \left( \frac{(n+p)!}{n!} \right)^2 (v(n)
v(n+p) - v(n)^2 v(n+p)^2 ) .
\eea
In view of (\ref{y9}) this behaves as
\be
{\rm Var}_w ({\cal O}) \sim \frac{\Gamma(2p+1)}{(-\log q)^{2p+1}}
\ee
and in particular
\be \label{y14}
\frac{{\rm Var}_w ({\cal O})}{(\langle {\cal O} \rangle_w)^2} \sim
\frac{\Gamma(2p+1)}{\Gamma(p+1)^2} (-\log q).
\ee
Since $\log q$ behaves as $1/\sqrt{\Delta}$=1/N, we find the same
behavior in the variance of the correlation function as we found
in the previous example. This again agrees with the general
expectation from Section~2.2  that the variance should
behave as $1/\log S$, with $S$ the entropy.

When does the variance become large? Superficially, the prefactor
in (\ref{y14}) becomes of order one for $p\sim \log N$. This is
not quite what we would expect, we would expect a breakdown for
$p\sim N^{\gamma}$ for some $\gamma$ rather than for $p\sim \log
N$. This may well be due to the various approximations that we
made in these estimates and we leave a more detailed study of this
to future work. Amongst other things, general correlations cannot
simply be computed in the matrix model, an integrable subsector of
the Yang-Mills theory, as we are doing here.    We can explore the
point where the variance becomes of order one also emerge when we
consider multi-trace operators that will probe $k$ different
harmonic oscillator levels at the same time. Then in computing the
variance as we did above, there are roughly $k^2$ possibilities
for special arrangements of the eigenvalues probed by the two
operators where at least two of them coincide. Thus we obtain an
enhancement of the variance by an extra factor of $k^2$,
suggesting a relative variance of order one when $k\sim N^{1/2}$.
More analysis is needed to define and analyze the universality of
correlation functions.

%%%%%%%%%%%%%%%%%%% Discussion %%%%%%%%%

\section{Discussion}

In this paper we have shown that very heavy pure states of quantum gravity can have an underlying quantum description as a ``foam'' with complex structure at the Planck scale, the details of which are invisible to almost all probes.
The ``foam'' might be more quantum mechanical or more classical depending upon the nature of the probe.
Whether one has a useful geometric description of the underlying microstate depends on the observable that one is measuring.
Specific classical geometries give the universal low-energy, long-wavelength descriptions of large classes of the underlying quantum states.
We showed how these classical descriptions are derived and how smooth underlying quantum dynamics can give rise to singular effective semiclassical geometries.
As an example, we explicitly constructed the singular spacetimes giving the universal low-energy description of almost all half-BPS states with a fixed mass.
The striking resemblance between the reasoning for the half-BPS states and the qualitative analysis of Schwarzschild black holes in Section 2 suggests that our exact results in the supersymmetric scenario give a generally correct qualitative understanding of  the non-supersymmetric case.

In the situations we have described, it is explicit that the low-energy classical description erases many crucial quantum mechanical details that are present at the Planck scale.
Information loss is thus implicit in the classical description of spacetime.
But when the detailed quantum mechanical wavefunction is incorporated there is no room for a breakdown of unitarity.
It has been suggested in \cite{juaneternal,hawkingrecent} that the information loss paradox for black holes is solved because the wavefunction for a black hole spacetime also has support on geometries that do not have a horizon.
However, it was also argued in \cite{eliezer,rabadan} that summing over geometries in this way is insufficient to restore information apparently lost in black holes.
We are proposing that the quantum mechanical wavefunction is indeed important, but sums over particular smooth and singular geometries using the Euclidean approach to quantum gravity will not suffice to restore unitarity.
The black hole in our picture is simply the universal effective low-energy description of a complex underlying state.
The loss of information lies precisely in the universality of the semiclassical description.\footnote{
It seems to us that recent attempts to ``see behind a horizon'' using
AdS/CFT \cite{horizon} will also not be able to identify the state
underlying a black hole precisely because these efforts never account
for the details of particular microstates. It would be interesting to
understand whether the structures in the complex time plane appearing
in these papers arise in some effective way from the underlying
``foams'' described in this paper.}

To see an actual horizon emerge in a low-energy effective description, one will have to consider states will less supersymmetry.
It may also be that interaction effects at finite coupling will become important.
Any star will collapse to a black hole if the Newton coupling is large
enough, indicating the appearance of horizons does depend on the coupling.
This is despite the fact that $g_s$ drops out of the expression for
the conformal dimension of a black hole operator. It will be important
to understand this issue and whether there is any phase transition as
a function of the coupling involved.   Nevertheless, it seems  that the heavy
states that we are describing know about gravity and don't just behave
like thermal gases. Probes of a standard thermal gas in a field theory
that have energies above the thermal scale pass through the gas
completely untouched. In our case, a probe of very high conformal
dimension interacts very strongly with any very heavy state simply
because there are more ways for the many fields in the state and probe operators to interact.
We suggest that this is an essentially gravitational phenomenon --- heavy objects
interact more strongly because, in gravity, the mass is the coupling constant.

Our results raise many questions including the following:
\begin{enumerate}
\item The limit shapes we have discussed give a ``mean field'' description of the microscopic physics.
The limit shape translates to a semiclassical geometry on the gravity side and gives universal correlation functions in the field theory because small-scale fluctuations have been integrated out.
In a sense, the limit shape is a ``master field''.
Fluctuations about the limit shape should be thought of in terms of continuum field theory.
It would be interesting to understand this structure more explicitly.
\item Can we assess correlation functions in greater detail to understand what information about the microstates of gravity can be extracted by different kinds of probes in both the half-BPS sector, and more generally without supersymmetry?
We expect that information about global charges is easy to determine and that detailed information about the nature of microstates will require carefully designed probes capable of making exceedingly fine measurements.
The elementary analysis of the Schwarzschild black hole suggests that
it is far easier to decide what a black hole is not rather than what it is.
Distinguishing the state would at least require determining all of the multipole moments.
How may we do this efficiently?
\item We have discussed a basis of half-BPS states related to single Young diagrams preserving $U(1)$ symmetry.
Generic half-BPS states, which are sums of these diagrams break this symmetry.
A particularly convenient method for describing the semiclassical limit of such states is in terms of coherent states.
This gives rise to a semiclassical picture of topological complex Planck scale droplets whose coherent sum gives rise to smooth geometries.
What can this teach us about the underlying quantum structure of gravity and the emergence of classical spacetimes?
\item While the $N$ fermions that appear in the description of half-BPS states have a $2N$-dimensional phase space, only the effective two-dimensional one-particle phase space makes an appearance in classical gravity.
What is the r\^ole of the full $N$-particle phase space and what does the full quantum system in the field theory teach us about boundary conditions at the singularity in the effective classical spacetime?
\item How do the lessons from the analysis of typical half-BPS states in $\AdS{5}\times S^5$ translate to other settings?   Can a similar picture of underlying ``foam'' and effective singular geometries be developed, for example, for the D1-D5 string that underlies most extremal black holes in string theory? Some efforts in this direction, using the techniques of Mathur and collaborators \cite{mathur}, will be reported in \cite{masaki1}. 
\item Can stringy corrections ``cloak'' the singular effective geometries describing our generic half-BPS states with a horizon, in a manner similar to \cite{cloaking}?
Does the area of the horizon match the entropy of the associated microstates?
\item What is the relation between our analysis of typical half-BPS states and the description of quantum foam in the topological string \cite{topological}?
In particular, there are strong similarities in the appearance of limit shapes of Young diagrams and topologically complex geometries containing random insertions of blown-up spheres in an otherwise smooth and locally flat manifold.
\item In Section~2 we gave arguments justifying our expectations that the conformal dimension of heavy operators $(\Delta\sim N^2)$ would not renormalize strongly.
It would be interesting to analyze this point precisely.
Along the same lines of extrapolating from weak to strong coupling, it would be interesting to understand better the r\^ole of (possible) phase transitions in the appearance of bulk horizons at strong coupling in the dual field theory formulation.
\item In principle, there is no upper bound to the dimension of operators in the Yang--Mills gauge theory.
What do typical states of conformal dimension greater than $N^2$ look like in gravity?
\item In \cite{llm} the phase space droplets giving rise to nonsingular spacetimes are incompressible Fermi liquids.   As a result there is an exact dictionary to the quantum Hall system \cite{beren, shahin} along with a $W_\infty$ algebra of area preserving diffeomorphisms.
The grayscale distribution in our analysis describes regions of the phase space plane where the lowest Landau level has filling fraction less than one.
It would be interesting to develop an understanding of this from the statistical mechanical point of view, especially with respect to possible relations to the fractional quantum Hall effect.
\end{enumerate}

We have proposed a picture in which the thermodynamic character of very heavy states in gravity arises because the effective description in geometry integrates over quantum details.
As is well known, inertial observers in empty de Sitter space and Rindler observers in empty flat space also see horizons with an entropy and a thermal character.
How might this arise?
From our experience in field theory, we are used to the decoupling theorem that says that ultraviolet physics decouples from the infrared.
As a result, integrating out the ultraviolet physics in a field theory vacuum state can lead to an infrared effective field theory in a pure vacuum state.
However, the decoupling theorem fails in theories with gravity.
It is tempting to speculate that, in this context, integrating ultraviolet details can lead to a low-energy mixed state because of entanglement between the infrared and ultraviolet parts of the wavefunction of spacetime.
Perhaps this entanglement and the resulting mixed nature of the low-energy description are more readily visible in some kinematic frames in which an observer effectively has large energies with respect to the natural Hamiltonian.

\section*{Acknowledgments}
We thank Ofer Aharony, David Berenstein, Micha Berkooz, Ramy
Brustein, Alex Buchel, Ben Craps, Bartek Czech, Gary Horowitz, Charlie Kane, Per
Kraus, Klaus Larjo, Tommy Levi, Gautam Mandal, Don Marolf, Liat Maoz, David Mateos, Rob
Myers, Asad Naqvi, Kyriakos Pappadodimas, Soojong Rey, Simon Ross,
Moshe Rozali, Shahin Sheikh-Jabbari, Masaki Shigemori, Nemani
Suryanarayana, and Steve Shenker for useful discussions.
We are as well grateful to Ben Craps and Tommy Levi for pointing out typographical errors in the first preprint version of this paper.
We are
also grateful to Taichiro Kugo for writing the young.sty and
wick.sty LaTeX packages. An essay based on some of the material in
this paper has appeared in \cite{GRF}. We thank the organizers of
the OCTS conference on black holes, the Uppsala string cosmology
workshop, the Fields Institute workshop on gravitational aspects
of string theory, the Euclid meeting at SISSA, and Strings 2005
for creating stimulating environments where significant progress
on this project was made. V.B.\ is particularly grateful to the
American University in Cairo for hospitality and logistical
support that made this paper possible. J.S.\ would like to thank
the Weizmann Institute of Science for kind hospitality during the
realization of this project. V.B.\ and J.S.\ are supported in part
by the DOE under grant DE-FG02-95ER40893, by the NSF under grant
PHY-0331728 and by an NSF Focused Research Grant DMS0139799. V.J.\
is supported by PPARC. JdB is supported in part by the FOM
foundation.

While this paper was being completed we became aware of several other
works on related topics \cite{davidrecent,shepard,worksinprogress1}.

\appendix

\section{Comments on correlation functions}

\subsection{Semiclassical distribution functions -- another derivation}

Given an arbitrary $N$-particle quantum state, when we integrate over
$N-1$ of the particles, we obtain a single-particle state described in
terms of a density matrix. If we consider this procedure among typical
states in the ensembles discussed in this paper, the average density
matrix $\rho_w$ that we would expect to obtain is
\be
\rho_w = \sum_{n \geq 0} c(n) |n\rangle \langle n |
\ee
For example, in the ensemble in which there is no constraint in the
number of columns of the Young diagram, the collection $\{c_n\}$ is
thermal, and precisely given by the Fermi-Dirac distribution
\be
c_{(n)} = \frac{q^{n-\mu}}{1+q^{n-\mu}}~.
\ee
for a suitable chemical potential $\mu$. This is a meaningful
statement since we do expect the probability of a single fermionic
excitation to be $n$ to equal the above distribution.

As has been argued along Section~4, the average density matrix
$\rho_w$ is related to the Husimi phase space distribution as
\be
u(p,q) = \sum_{n \geq 0 } c(n) {\rm Hu}_n(z)
\ee
where as before $z=(p+iq)/\sqrt{2\hbar}$.
In the semiclassical limit, where the set of $\{c_n\}$ varies very
slowly as a function of $n$, our considerations suggest that
\be
u(p,q) = \frac{1}{2\pi\hbar} c(|z|^2)~.
\ee
To show this in more detail, assume that
\be
c(n) = \sum_s a_s e^{b_s n}
\ee
with $b_s \ll 1$. Then
\bea
u(p,q)&  =&  \frac{1}{2\pi \hbar} \sum_{s,n} a_s e^{b_s n} \frac{|z|^{2n}}{n!} e^{-|z|^2}
\nonumber \\
& = & \frac{1}{2\pi \hbar} \sum_{s} a_s e^{|z|^2(e^{b_s}-1)} \nonumber \\
& \sim & \frac{1}{2\pi \hbar} \sum_{s} a_s e^{b_s |z|^2} \nonumber \\
& = & \frac{1}{2\pi\hbar} c(|z|^2).
\eea
Thus, at least in the semiclassical limit, the phase space
distribution function $u(p,q)$ should equal the Fermi-Dirac
distribution, an statement that we also derived from the slope in the
limit curve of the limit Young diagram through (\ref{proposal2}).

\subsection{Single trace correlation functions}
We are interested in computing the variance of a correlation function in an ensemble of partitions:
\be
\sigma^2 = \sum_{\lambda} w(\lambda) \langle \lambda | F | \lambda \rangle \langle \lambda
 | F | \lambda \rangle- \left( \sum_{\lambda} w(\lambda) \langle \lambda | F| \lambda \rangle \right)^2 .
\ee
In this expression $\lambda$ corresponds to a partition and we will write $n\in \lambda$ to indicate that
harmonic oscillator level $n$ is occupied when the fermions are in the state corresponding
the partition $\lambda$. As in Section~5, let $F$ be a rotationally invariant sum of one-fermion operators.  Then
\be
\langle \lambda | F| \lambda \rangle = \sum_{n\in\lambda} \langle n |F|n \rangle.
\ee
The probability that $n\in \lambda$ is
\be v(n) =  \sum_{\lambda,n\in\lambda} w(\lambda)
\ee
and since we assume that we work in a grand canonical ensemble where the probability that $n$ and $m$ are
occupied is simply $v(n) v(m)$, we can rewrite $\sigma^2$ as
\bea
\sigma^2 & = & \sum_{\lambda} \sum_{n\in\lambda} \sum_{m\in\lambda} w(\lambda)
\langle n|F|n \rangle \langle m|F|m \rangle \nonumber \\
& & - \sum_{\lambda} \sum_{\mu}  \sum_{n\in\lambda} \sum_{m\in\mu} w(\lambda) w(\mu)
\langle n|F|n \rangle \langle m|F|m \rangle \nonumber \\
& = & \sum_{n} v(n)(\langle n|F|n \rangle)^2 + \sum_{m\neq n} v(m) v(n)
\langle n|F|n \rangle \langle m|F|m \rangle \nonumber \\
& & - \sum_n v(n)^2 (\langle n|F|n \rangle)^2 - \sum_{m\neq n} v(m) v(n)
\langle n|F|n \rangle \langle m|F|m \rangle \nonumber \\
& = & \sum_n (v(n)-v(n)^2) (\langle n | F | n \rangle)^2
\nonumber \\
& = & \sum_n (v(n)-v(n)^2) \langle n | F^2 |n \rangle.
\eea
In the last line we used the rotational invariance of $F$.  This
is simply the expectation value of $F^2$ in the  density matrix
$\sum_n (v(n)-v(n)^2)|n\rangle\langle n |$.   As we described in
Section~A.1, in the semiclassical limit this becomes an integral over
the $p,q$ plane of $u(p,q)-u(p,q)^2$ times $F(p,q)^2$. Notice that
for fermionic statistics the variance in the occupation number
$\sigma^2 =  \langle n^2\rangle - \langle n\rangle^2= \langle
n\rangle -\langle n\rangle^2$ since $n$ only takes the values $0$
and $1$. This implies that $u-u^2$ can be interpreted as the
variance in the fermion occupation number, and the expression here
is in fact equivalent to the one given in (\ref{sigsig}). Thus,the
variance in the expectation value of $F$ is the integral of $F^2$
against the variance in the fermion occupation number. In Section~5
we derived a closely related expression relating the variance in
$F$ to the variance in the distribution function on phase space.
Following the same reasoning as in that section, one can argue
that the deviations from universality for delocalized operators
goes as $e^{-N}$. This meets our expectations since in the
ensembles of interest to us, the entropy scales like $N$.

\end{document}